\PassOptionsToPackage{square,comma,numbers,sort&compress,super}{natbib}
\documentclass[prx,aps,preprint,a4paper,longbibliography]{revtex4-1} 
\usepackage{graphicx}
\usepackage[version=3]{mhchem}
\usepackage{dcolumn}
\usepackage{multirow}
\usepackage{bm}
\usepackage{subfigure}
\usepackage{amsmath}
\usepackage{amssymb}
\usepackage{epsfig}
\usepackage{epstopdf}
\begin{document}
\title{Origin of superconductivity and giant phonon softening in TlInTe$_2$ under pressure}
\author{Sorb Yesudhas$^{1*}$, N. Yedukondalu$^{2*,3}$, Manoj K. Jana$^4$, Jianbo Zhang$^1$, Jie Huang$^1$, Bijuan Chen$^1$, Hongshang Deng$^1$, Raimundas Sereika$^{1,5}$, Hong Xiao$^1$, Stanislav Sinogeikin$^6$, Curtis Kenney-Benson$^7$, Kanishka Biswas$^4$, John B. Parise$^{2,8}$, Yang Ding$^{1*}$, Ho-kwang Mao$^1$}
\affiliation{$^1$Center for High-Pressure Science $\&$ Technology Advanced Research, Beijing 100094, P.R. China \\
 $^2$Department of Geosciences, Center for Materials by Design, and Institute for Advanced Computational Science, State University of New York, Stony Brook, New York 11794-2100, USA \\
 $^3$Rajiv Gandhi University of Knowledge Technologies, Basar, Telangana-504107, India \\
 $^4$New Chemistry Unit, Jawaharlal Nehru Centre for Advanced Scientific Research (JNCASR), Bangalore, India \\
  $^5$Vytautas Magnus University, K. Donelai\v{c}io Street 58, Kaunas 44248, Lithuania \\
   $^6$DAC Tools, Custom Equipment for High-Pressure Research, Naperville, IL 60565-2925, USA \\
  $^7$HPCAT, X-ray Science Division, Advanced Photon Source, Argonne National Laboratory, Lemont IL, 60439, USA \\
   $^8$National Synchrotron Light Source II, Brookhaven National Laboratory, Upton, New York 11973, USA \\
 }
\date{\today}
\begin{abstract}
Analogous to 2D layered transition metal dichalcogenides, the TlSe family of 1D chain materials with Zintl-type structure exhibits exotic phenomena under high-pressure. In the present work, we have systematically investigated the high-pressure behavior of TlInTe$_2$ using Raman spectroscopy, synchrotron X-ray diffraction, and transport measurements, in combination with crystal structure prediction (CSP) based on the evolutionary approach and first principles calculations. We found that TlInTe$_2$ undergoes a pressure driven semiconductor to semimetal transition at 4 GPa, followed by a superconducting transition at 5.7 GPa (with T$_c$ = 3.8 K) induced by a Lifshitz transition. The Lifshitz transition is initiated by the appearance of new electron pockets on the Fermi surface, which evolve with pressure and connect to the adjacent electron pockets forming an umbrella shaped Fermi surface at the top and bottom of the Brillouin zone. An unusual giant phonon softening (A$_g$ mode) concomitant with a V-shaped T$_c$ behavior appears at 10-12 GPa as a result of the interaction of optical phonons with the conduction electrons, resulting in Fano line shaped asymmetry in A$_g$ mode. A prominent T$_c$ anomaly  concurrent with the A$_g$ mode softening at 19-20 GPa is correlated to the semimetal to metal transition. The CSP calculations reveal that these transitions are not accompanied by any structural phase transitions up to the maximum pressure achieved, 33.5 GPa. Our findings on TlInTe$_2$ open up a new platform to study a plethora of unexplored high pressure novel phenomena in TlSe family induced by Lifshitz transition (electronic driven), phonon softening and electron-phonon coupling.
\end{abstract}

\maketitle
\section{Introduction}
Soon after the discovery of non-trivial topological quantum states in chalcogen based semiconductors with narrow band gaps and strong spin-orbit couplings (SOC), the two dimensional (2D) layered semiconductors opened up a new avenue to scrutinize novel pressure induced phenomena, such as quantum phase transitions, topological superconductors, charge density waves, structural phase transitions, Lifshitz transitions etc. \cite{zhang2011pressure,kung2019observation,zhu2011substitutional,sorb2015pressure} In contrast to layered 2D materials, one dimensional (1D) chain materials such as TlInTe$_2$, TlInSe$_2$ and TlGaTe$_2$ are relatively under explored from the perspective of high pressure research although they exhibit astonishing properties at ambient pressure.\cite{panich2008electronic, dutta2019ultralow} For instance, TlInSe$_2$ exhibits exceptionally high thermoelectric properties, which is correlated to the formation of an incommensurate superlattice.\cite{mamedov2006super} TlInTe$_2$ exhibits ultra-low thermal conductivity ($\textless$ 0.5 W/mK in the temperature range 300-673 K)\cite{jana2017intrinsic} owing to the spatial fluctuation of Tl$^{1+}$ cation inside the polyhedral framework of a sublattice formed by Tl$^{1+}$ and Te$^{2-}$ ions.

The Zintl type chain material, TlInTe$_2$ crystallizes in tetragonal structure having space group, $I/4mcm$ (B37) with Z = 4, which is iso-structural to its parent TlSe compound.\cite{ellialtiouglu2004electronic} The unit cell of TlInTe$_2$ has two sublattices, viz. In$^{3+}$Te$_4^{2-}$ tetrahedral and Tl$^{1+}$Te$_8^{2-}$ polyhedral (with eight vertices) sublattices. In In$^{3+}$Te$_4^{2-}$ sublattice, four Te$^{2-}$ anions are tetrahedrally bonded to an In$^{3+}$ cation, with the corner sharing of Te atoms along $c$-axis forming an anionic chain. Similarly, the Tl$^{1+}$ cation is loosely bonded to eight Te$^{2-}$ ions in the Tl$^{1+}$Te$_8^{2-}$ sublattice, forming a cationic chain along $c$-axis, where the Tl$^{1+}$ ions spatially fluctuate inside the square anti-prismatic (Thomson cube) cage.\cite{jana2017intrinsic} The Te$^{2-}$ anions in the anionic chains are interconnected via covalent bonding (intrachain bonding) whereas the anionic and cationic chains are connected to each other by weak ionic bonding (interchain bonding). Thus the presence of two different bonding schemes, such as intrachain and interchain bonding in TlInTe$_2$ combined with its narrow band gap and strong SOC, provide a platform to explore a plethora of pressure induced exotic phenomena including semiconductor to semimetal transition, structural phase transitions, superconductivity, non-trivial topological states, Lifshitz transition, and an incommensurate phase formation (due to the spatial fluctuation of Tl$^{1+}$ cation inside the Thomson cube). The anisotropic bonding scheme in TlInTe$_2$ is anticipated to allow different compression rates along $a$ and $c$ directions, which may leads to bonding changes, gradual symmetry lowering structural transitions, anisotropic bond length changes etc.\cite{sorb2013high,Yedu2017} Astonishingly, InTe is iso-structural to TlInTe$_2$, which exhibits superconductivity. \cite{geller1964superconductivity} Due to an analogous structural relationship between InTe and TlInTe$_2$, we anticipate that TlInTe$_2$ would also exhibit superconductivity under moderate pressure. In addition, the mechanism giving rise to the origin of superconductivity in the TlSe family has not been explored yet. Strikingly, the monolayer of TlSe (S.G:$I4/mcm$) is predicted to be a topological insulator under uniaxial compression\cite{niu2015two} while InTe shows a band inversion under hydrostatic pressure.\cite{rajaji2018pressure}. The electronic structure and Fermi surface topology changes under high pressure can provide insights into band inversion, topological and Lifshitz phase transitions.\cite{Li2018,Bianconi2015,Jarlborg2016,Durajski2017} Moreover, the role of electron-phonon interactions, the evolution of superconducting transition temperature (T$_c$) with pressure, and the origin of superconductivity are yet to be investigated on this class of materials.

Preliminary high pressure Raman studies on TlInTe$_2$ up to $\sim$17 GPa suggested the emergence of a sluggish phase transition at 7 GPa due to the breaking or rearrangement of chains, which appeared to coexist with the parent tetragonal phase (B37) up to 17 GPa.\cite{ves1990high} Because of a lack of high pressure X-ray diffraction studies, the existence of a structural transition proposed at 7 GPa is still undetected. Similarly, room temperature resistivity measurements carried out on TlInTe$_2$ along $c$-direction exhibited continuous metallization up to 7 GPa upon cooling to liquid nitrogen temperature.\cite{rabinal1993effect} Since there is no systematic high pressure transport measurements on TlInTe$_2$, in this work, we investigate high pressure structural, vibrational and superconducting behavior of TlInTe$_2$ using high pressure Raman spectroscopy, synchrotron X-ray diffraction, and transport measurements combined with CSP using USPEX and first principles calculations based on density functional theory.

\section{Experimental details}
A detailed synthesis procedure of TlInTe$_2$ powder samples are discussed elsewhere. \cite{jana2017intrinsic} A symmetric type diamond anvil cell (DAC) with culet size of 300 $\mu$m was used for both high pressure synchrotron X-ray diffraction (XRD) and Raman measurements. A stainless-steel gasket was preindented to a thickness of 40 $\mu$m and a hole was drilled at its center to a diameter of 150 $\mu$m using laser drilling machine at High-Pressure Collaborative Access Team (HPCAT). A thin pellet of size 40 $\mu$m was loaded into the sample chamber. High pressure synchrotron XRD studies were carried out at 16-BM-D beamline at HPCAT, Advanced Photon Source. The hydrostatic condition was achieved by using Ne gas as pressure transmitting medium (PTM). The spot size of the X-ray beam used was $\sim$5 $\mu$m and the X-ray beams diffracted from the sample was detected by an imaging plate detector. The 2D XRD image was converted to a 1D pattern by using Fit2D software\cite{hammersley1996two} and the XRD data were fitted by using Fullprof Suite software.\cite{rodriguez1993recent} The pressure was determined by ruby fluorescence method.\cite{mao1976high}. Raman measurements were carried out on a Renishaw inVia Raman spectrometer and the sample was excited by using a green laser ($\lambda$ = 532 nm). The Ne gas was used as PTM for Raman measurements. The Raman data were collected using a CCD detector.

The transport measurements up to 3.9 GPa were carried out with Janis ST-500 continuous flow cryostat at to the temperatures down to 12 K and above 3.9 GPa, it was carried out using Quantum Design physical property measurement system (PPMS). High pressure was achieved by a Cu-Be diamond anvil cell and the resistance was measured by four probe method. Boron nitride powder was used to form an insulating gasket for the sample and Pt wires were used as electrical leads. No pressure transmitting medium was used for the transport measurements.

\section{Computational details}
First principles crystal structure prediction calculations were carried out using Universal Structure Predictor: Evolutionary Xtallography (USPEX) code based on evolutionary approach\cite{Oganov1,Oganov2,Oganov3}. We performed an extensive crystal structure search with fixed composition (TlInTe$_2$) at distinct pressures such as $\sim$ 0, 10, 30, and 50 GPa with 2, 4 and 8 formula units per primitive cell. The first generation with 150 structures is randomly generated and the succeeding 44 generations with a population size of 50 were obtained by applying heredity (50$\%$), random (20$\%$), soft mutation (20$\%$), and lattice mutation (10$\%$) operators until the best structure remains invariant up to 20 generations. The first principles calculations were performed within the framework of density functional theory (DFT). To obtain the global minimum energy structures, the structural optimization with projector-augmented plane-wave (PAW) potentials\cite{Joubert1999}, within the generalized gradient approximation of Perdew-Burke-Ernzerhof (PBE) parametrization\cite{Perdew1996} as implemented in the Vienna Ab-initio Simulation Package (VASP).\cite{Kresse1996} The PAW potentials with $5s^25p^65d^{10}6s^26p^1$, $4s^25s^24p^64d^{10}5p^1$ and $4s^25s^24p^64d^{10}5p^4$ electrons are treated as valence states for Tl, In and Te respectively. A kinetic energy cutoff of 530 eV was used for the plane wave basis set expansion and also 2$\pi$ $\times$ 0.024 $\AA^{-1}$ k-spacing has been chosen to sample the Brillouin zone.

Electronic structure and Fermi surface topology at high pressure were calculated using Tran-Blaha modified Becke Johnson (TB-mBJ) potential\cite{Tran2009} implemented through WIEN2k package.\cite{Blaha2002}  To achieve energy eigen value  convergence, wave functions in the interstitial region were expanded in plane waves with cutoff, K$_{max}$ = 7/RMT, where RMT is the smallest atomic sphere radius and K$_{max}$ denotes the magnitude of largest k vector in plane wave expansion, while the charge density was Fourier expanded up to G$_{max}$ = 14. The muffin-tin radii were assumed to be 2.6 Bohr for Tl and 2.5 Bohr for In, Te atoms.

\section{Results and discussions}
\section{High pressure Raman spectroscopy}
\begin{figure}
\centering
\subfigure[]{\includegraphics[width=3.1in,height=2.8in]{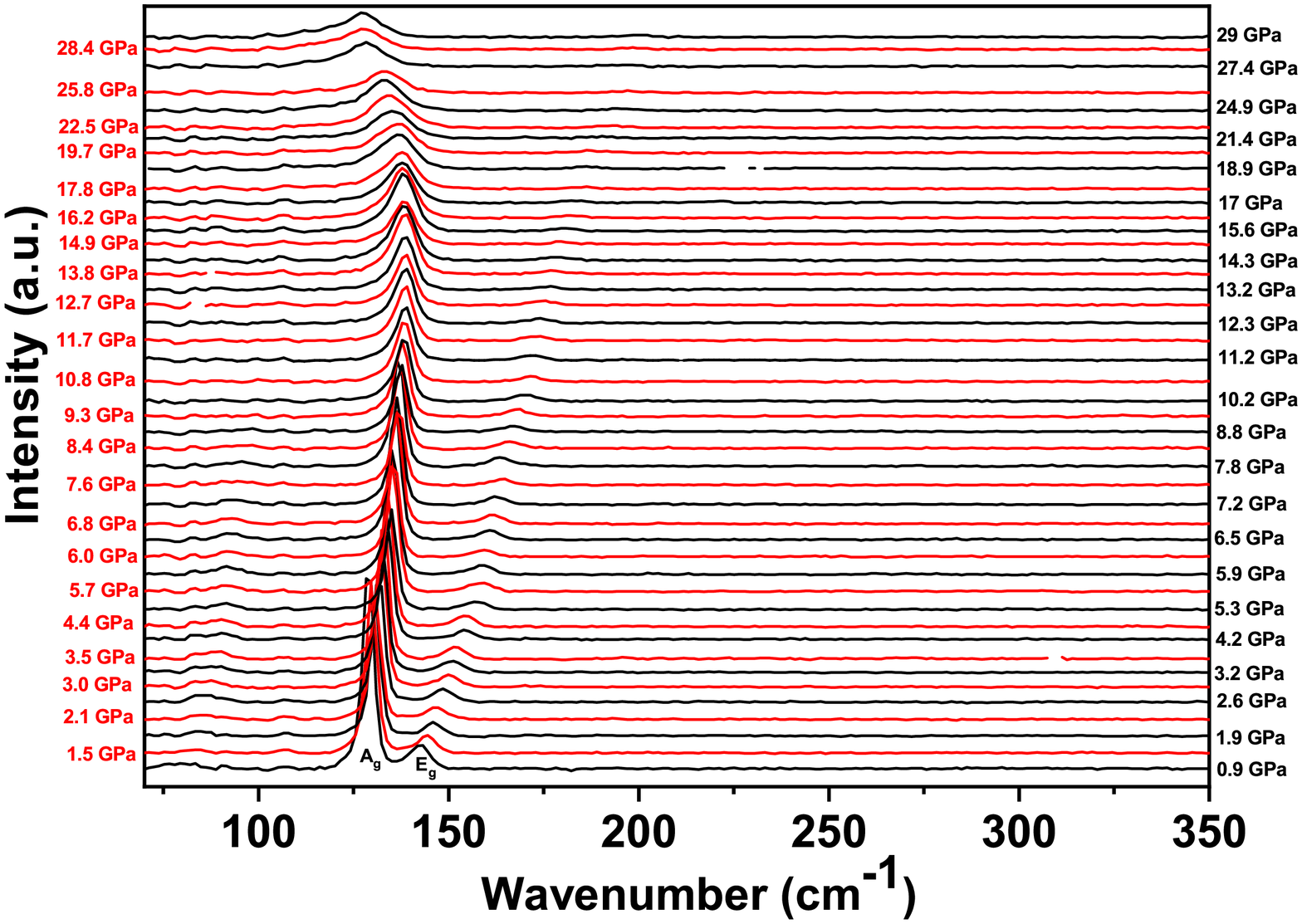}} 
\subfigure[]{\includegraphics[width=3.1in,height=2.8in]{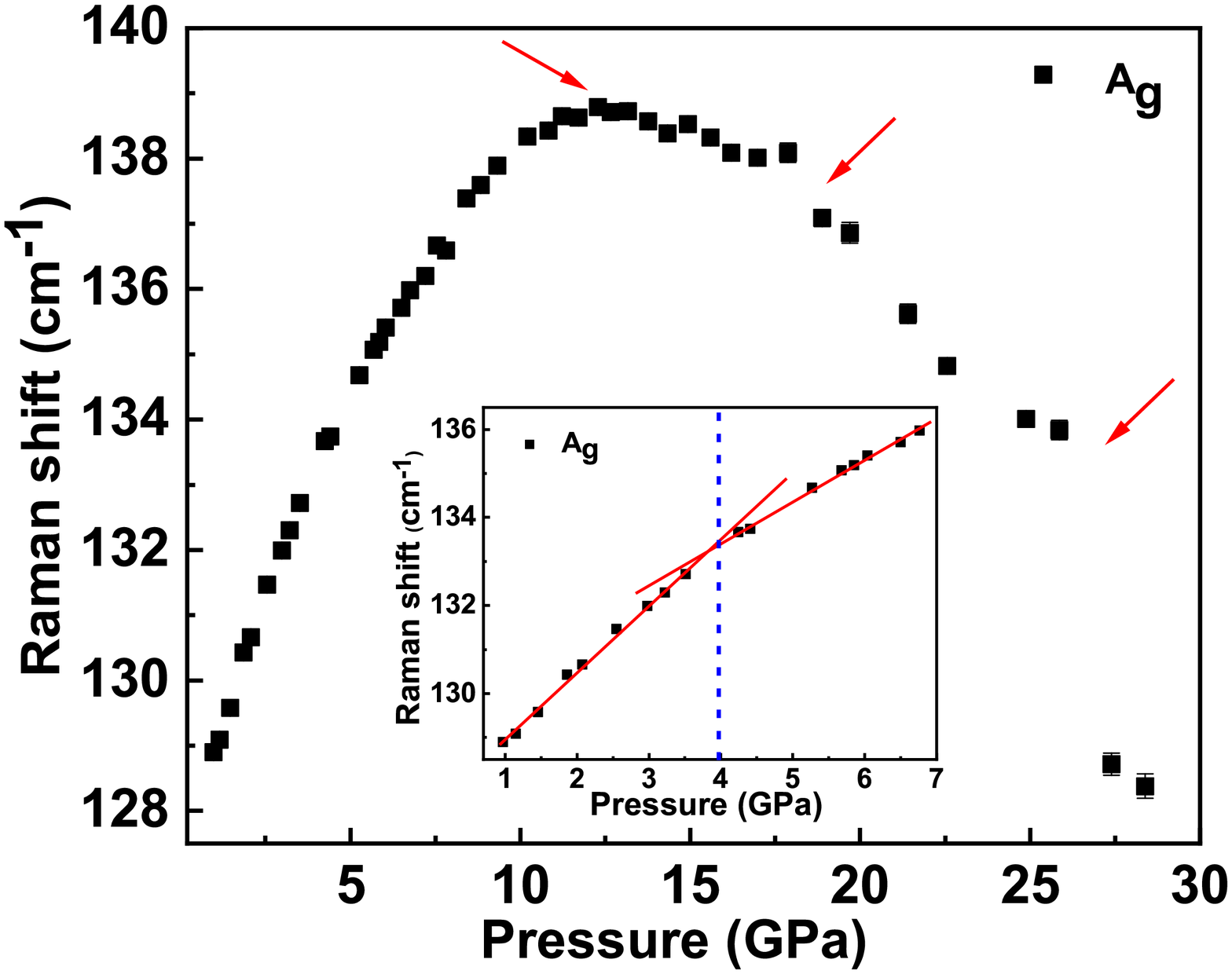}} 
\subfigure[]{\includegraphics[width=3.1in,height=2.8in]{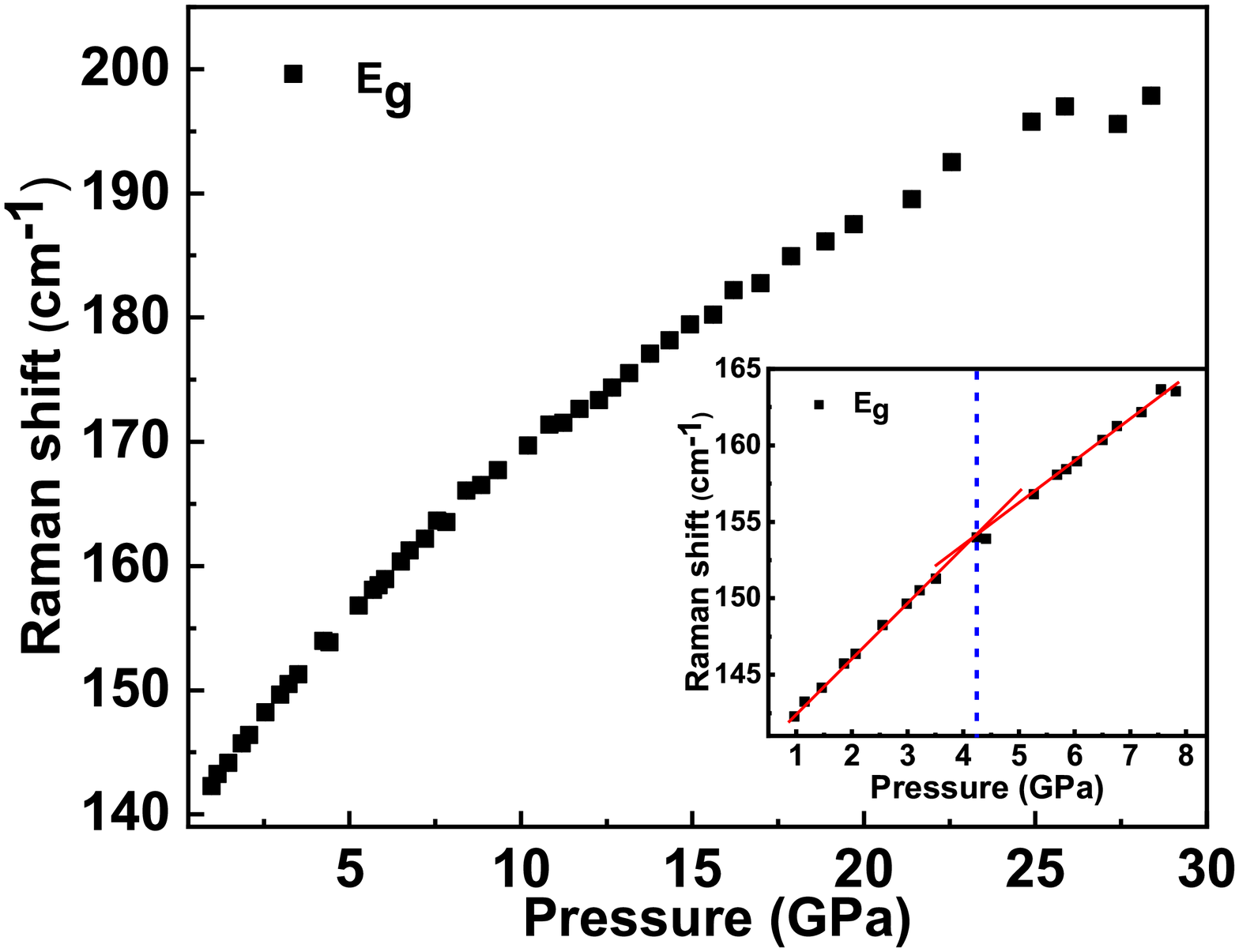}}
\subfigure[]{\includegraphics[width=3.1in,height=2.8in]{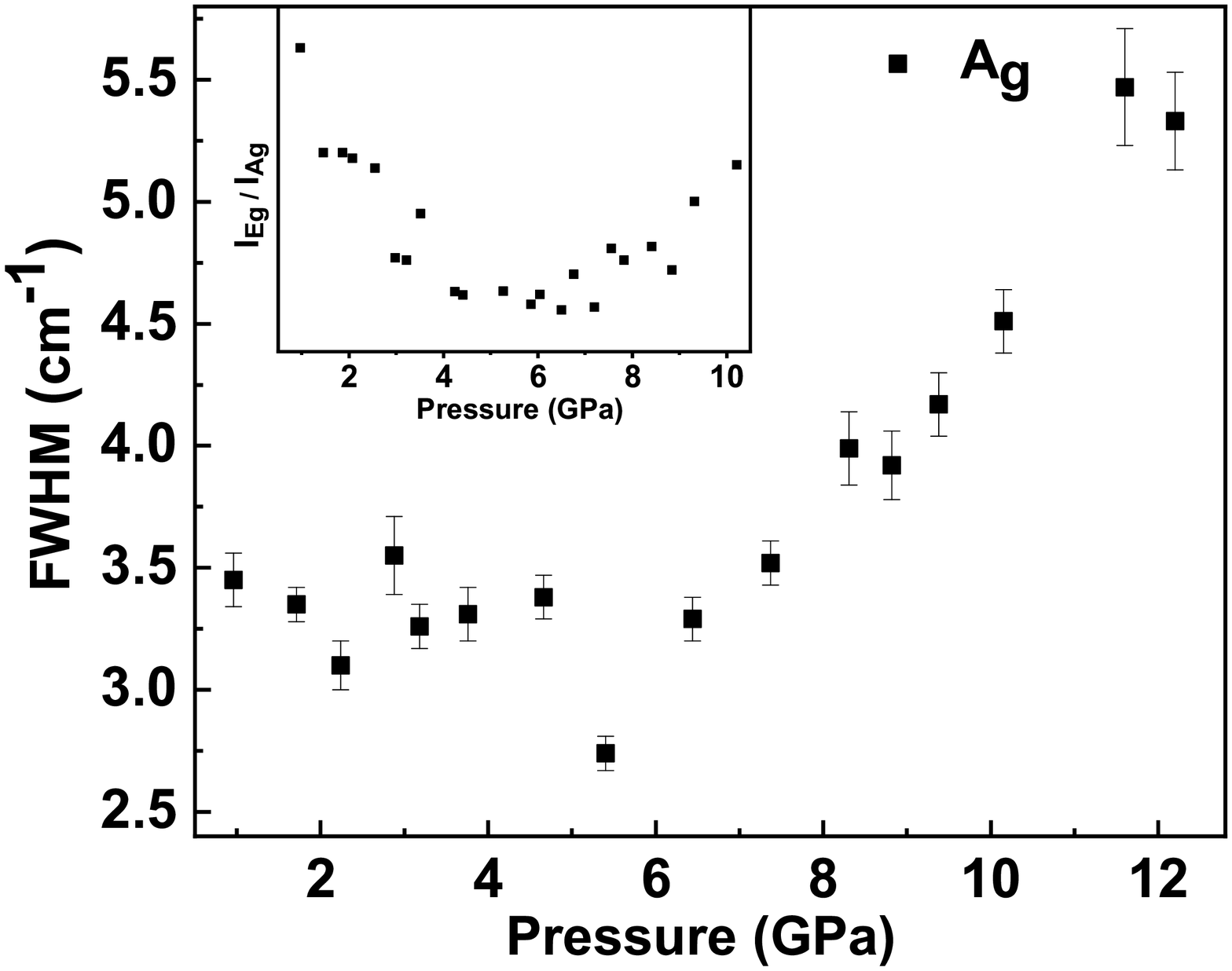}}
\caption{(a) Raman spectra of TlInTe$_2$ for various pressures. The successive Raman spectra are represented as black and red colors. (b) and (c) Pressure dependence of A$_g$ and E$_g$ Raman modes. Softening of Raman modes at 4 GPa are shown as inset and the dashed blue lines indicate transition pressure. The red lines represent guidance to the eyes. The slanted red arrows to indicate different transitions. The size of the errors is smaller than the size of the symbols. (d) Pressure versus FWHM (cm$^{-1}$) of A$_g$ mode with error bars. Pressure versus FWHM (cm$^{-1}$) of E$_g$ mode with error bars. The intensity ratio of E$_g$ and A$_g$ modes is shown as an inset.}
\label{fig:Raman}
\end{figure}

The primitive cell of TlInTe$_2$ has eight atoms, results in 24 (3N; where N is number atoms per primitive cell) phonon modes at the $\Gamma$ point of the Brillouin zone and the factor group analysis yields:
\begin{center}
$\Gamma_{vib}$ = A$_{1g}$ $\oplus$ 2A$_{2g}$ $\oplus$ B$_{1g}$ $\oplus$ 2B$_{2g}$ $\oplus$ 3E$_g$ $\oplus$ B$_{1u}$ $\oplus$ 3A$_{2u}$ $\oplus$ 4E$_u$
\end{center}
where the subscripts 'g' and 'u' denote gerade and ungerade modes,  respectively. Out of 24 vibrational (3 acoustic and 3N-3=21 optical) modes, seven are Raman active and five are IR active. Since our Raman spectrometer can detect only frequencies above 100 cm$^{-1}$, we have observed only two Raman active modes, viz. A$_g$ = 127.8  (127)\cite{gasanly1980vibrational} cm$^{-1}$ and E$_g$ = 139  (138)\cite{gasanly1980vibrational} cm$^{-1}$ of TlInTe$_2$ at ambient pressure, where the A$_g$ mode represents the opposite motion of adjacent Te$^{2-}$ anions in the In$^{3+}$Te$_4^{2-}$ tetrahedra and the E$_g$ mode represents the vibration of Te$^{2-}$ ions along the chain (XZ plane) or the motion of In$^{3+}$ cation along the XY plane as illustrated in Fig. S1a.\cite{gasanly1980vibrational,ves1990high} The Raman modes observed in this work are in excellent agreement with the previous Raman measurements\cite{gasanly1980vibrational} as well as with the recent theoretical calculations (A$_g$ = 128.2 and E$_g$ = 138.8 cm$^{-1}$).\cite{Minghui2019}. Raman spectra of TlInTe$_2$ at different pressures are as shown in Fig. \ref{fig:Raman}a and the pressure dependence of Raman modes, A$_g$ and E$_g$ are plotted in Fig. \ref{fig:Raman} b $\&$ c. It is evident from Fig. \ref{fig:Raman}a that the intensity of these Raman modes decreases with pressure. The A$_g$ mode exhibits a distinct softening whereas the E$_g$ mode shows a subtle softening at 4 GPa as shown in the inset of Fig. \ref{fig:Raman}b $\&$ c. Moreover, the intensity ratio of E$_g$ and A$_g$ modes exhibits a significant change at 4 GPa (Fig. \ref{fig:Raman}d). In order to further comprehend the high pressure behavior of Raman modes, pressure coefficients and mode Gr\"uneisen parameters are calculated by fitting the pressure-frequency data of phonon modes up to 4 GPa to a straight line. The calculated pressure coefficients of A$_g$ and E$_g$ modes are as shown in Table S1 and are closely comparable to the previously  reported values.\cite{ves1990high} The relative stiffness of the A$_g$ mode, with a pressure coefficient of 1.51 cm$^{-1}$/GPa compared to that of E$_g$ mode, 3.54 cm$^{-1}$/GPa, suggests a disparity in the bonding behavior of intra and inter chains of TlInTe$_2$ [see Table S1]. The bonding behavior and various forces acting on a crystal can be inferred from mode Gr\"uneisen parameters ($\gamma$), which are calculated from the above linear fit by using the following relationship:
\begin{equation}
\gamma(\omega) = \left(\frac{B_0}{\omega_0}\right)\frac{d\omega}{dp}\Big\lvert_{p=0}
\end{equation}

where $\omega_0$ is the frequency and B$_0$ is the bulk modulus at ambient pressure. The calculated B$_0$ of TlInTe$_2$ is 19.4 GPa (see Fig. \ref{fig:XRD}c). However, the calculated $\gamma$ values of A$_g$ ($\gamma$ = 0.23) and E$_g$ ($\gamma$ = 0.5) modes are much smaller compared to the reported values; A$_g$ ($\gamma$ = 0.81) and E$_g$ ($\gamma$ = 1.6) \cite{ves1990high} [see Table S1], which is due to large bulk modulus value (B$_0$ = 58.3 GPa; $\sim$ 3 times larger than the one obtained in this work) used for calculating the $\gamma$ values.\cite{ves1990high} It is interesting that the pressure dependence of the full width at half maximum (FWHM) of the A$_g$ mode is constant up to 6 GPa, while above this pressure, the phonon line width exhibits a dramatic increase [Fig. \ref{fig:Raman}d]. Similarly, the intensity ratio of E$_g$ and A$_g$ modes exhibits a minimum around 6 GPa [see the inset of Fig. \ref{fig:Raman}d]. The phonon life time change (FWHM (cm$^{-1}$)) of A$_g$ mode and a minimum in pressure versus intensity ratio of E$_g$ and A$_g$ modes hint at another transition at 6 GPa [Fig. \ref{fig:Raman}d]. Upon further compression, we observe major changes in the frequency of A$_g$ phonon mode as shown in Fig. \ref{fig:Raman}b. For instance, an abnormal frequency reversal (softening) of the A$_g$ mode, and a subtle softening of E$_g$ mode, are observed around 10-12 GPa. The anomalies at 19 GPa and 25 GPa are indicated by softening and discontinuity (collapse) of the frequency of A$_g$ mode, respectively. Since Raman spectra (Fig. \ref{fig:Raman}) do not provide any explicit evidence of either appearance or disappearance of phonon modes up to the maximum pressure achieved (29 GPa). This suggests that the origin of these transitions are unrelated to structural phase transitions. A more detailed investigation of the above transitions are discussed in the subsequent sections with reference to high pressure synchrotron XRD, transport measurements, crystal structure predictions and first principles calculations.

\subsection{High pressure synchrotron X-ray diffraction and crystal structure prediction}
\begin{figure}
\centering
\subfigure[]{\includegraphics[width=3.1in,height=2.8in]{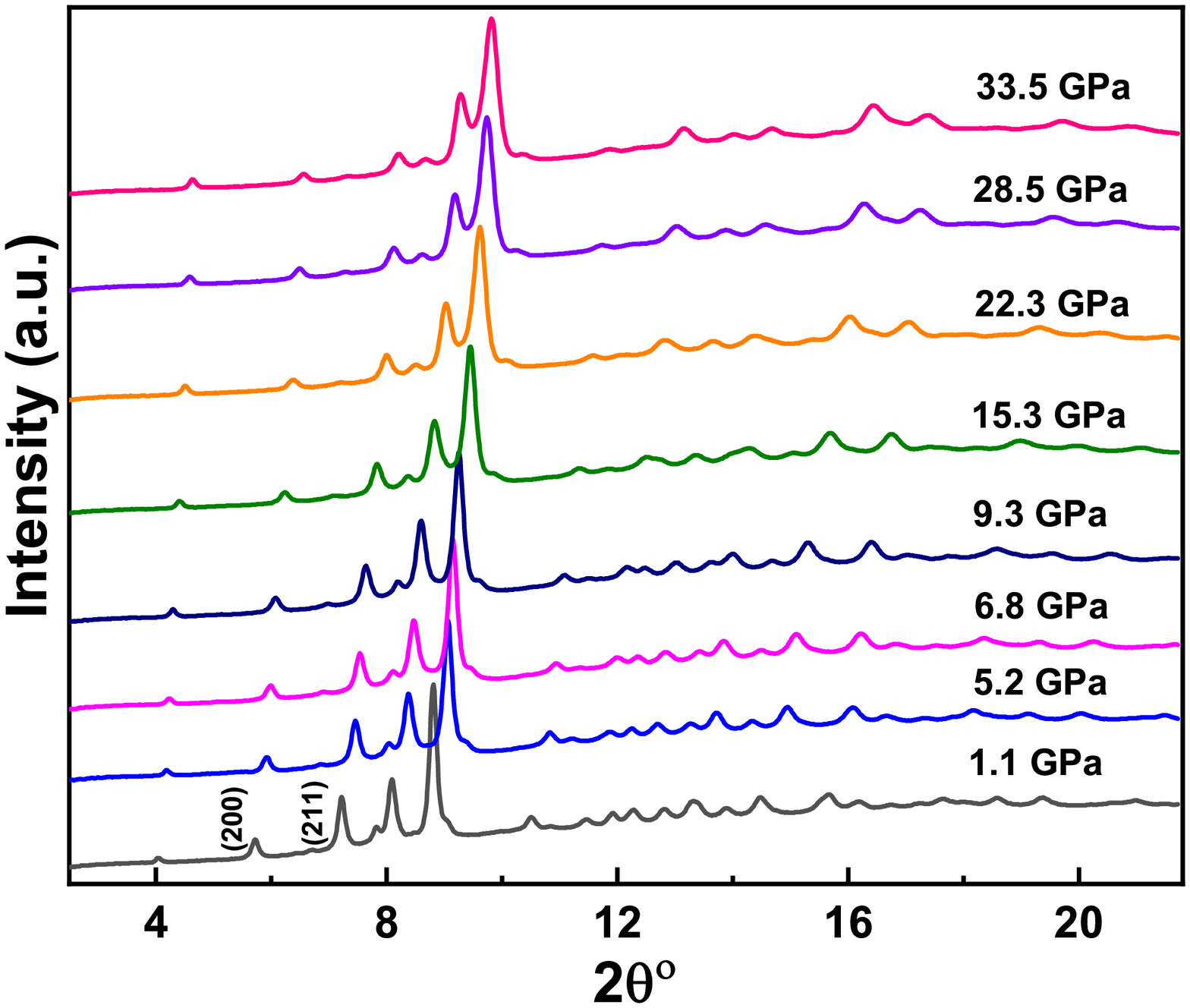}} \hspace{2mm}
\subfigure[]{\includegraphics[width=3.1in,height=2.8in]{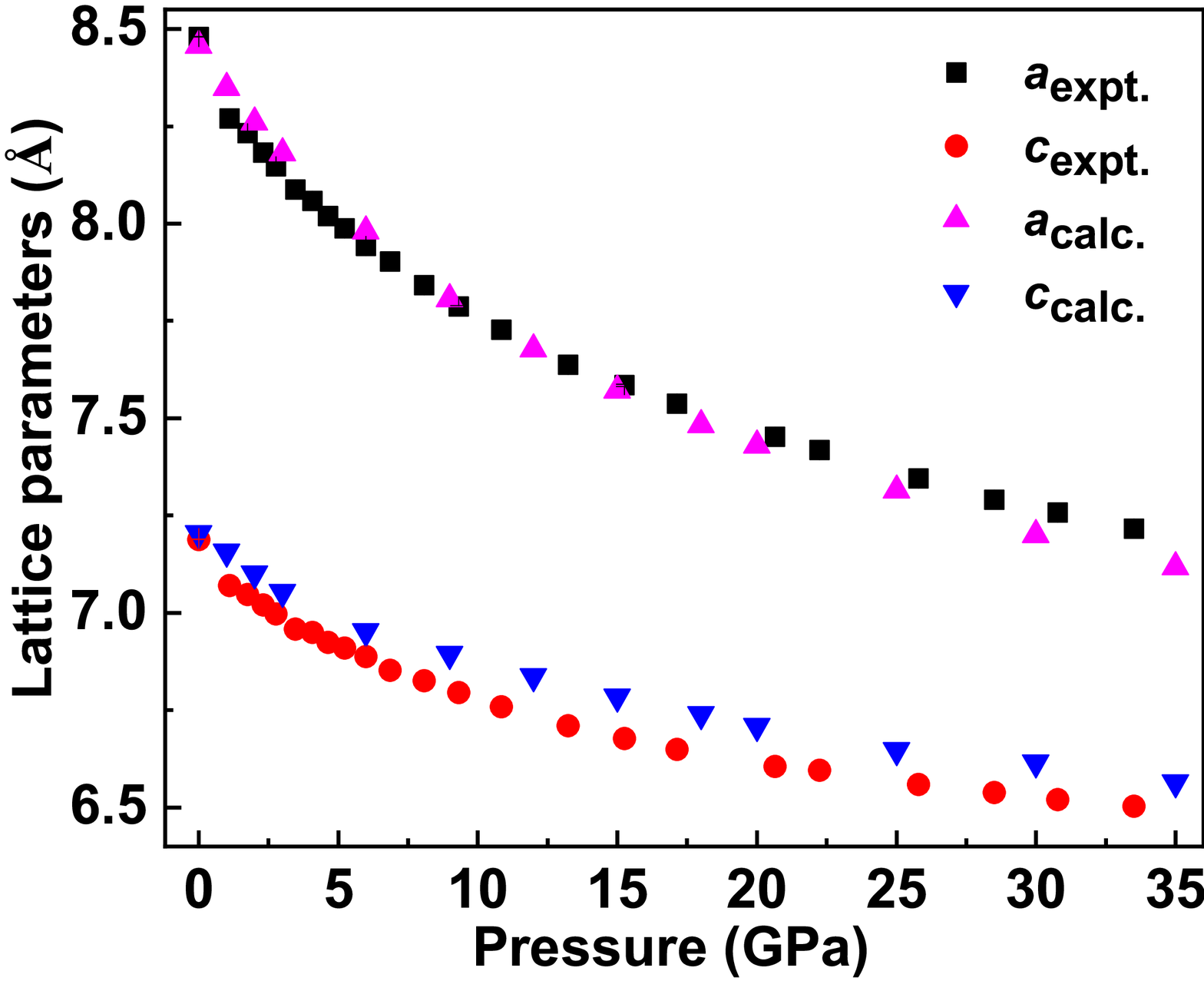}} \vspace{0.1 in}
\subfigure[]{\includegraphics[width=3.2in,height=2.8in]{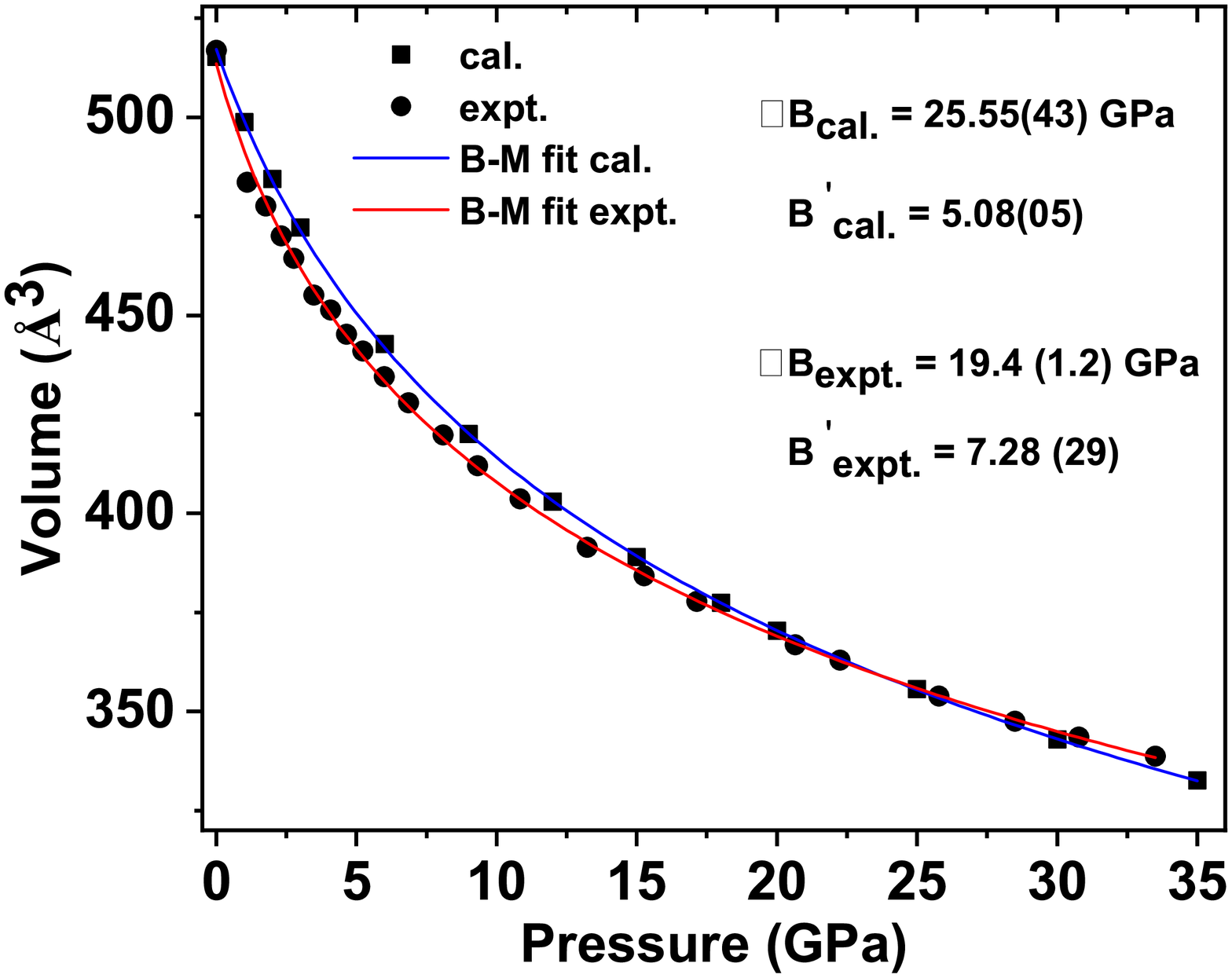}} \hspace{2mm}
\subfigure[]{\includegraphics[width=3.1in,height=2.8in]{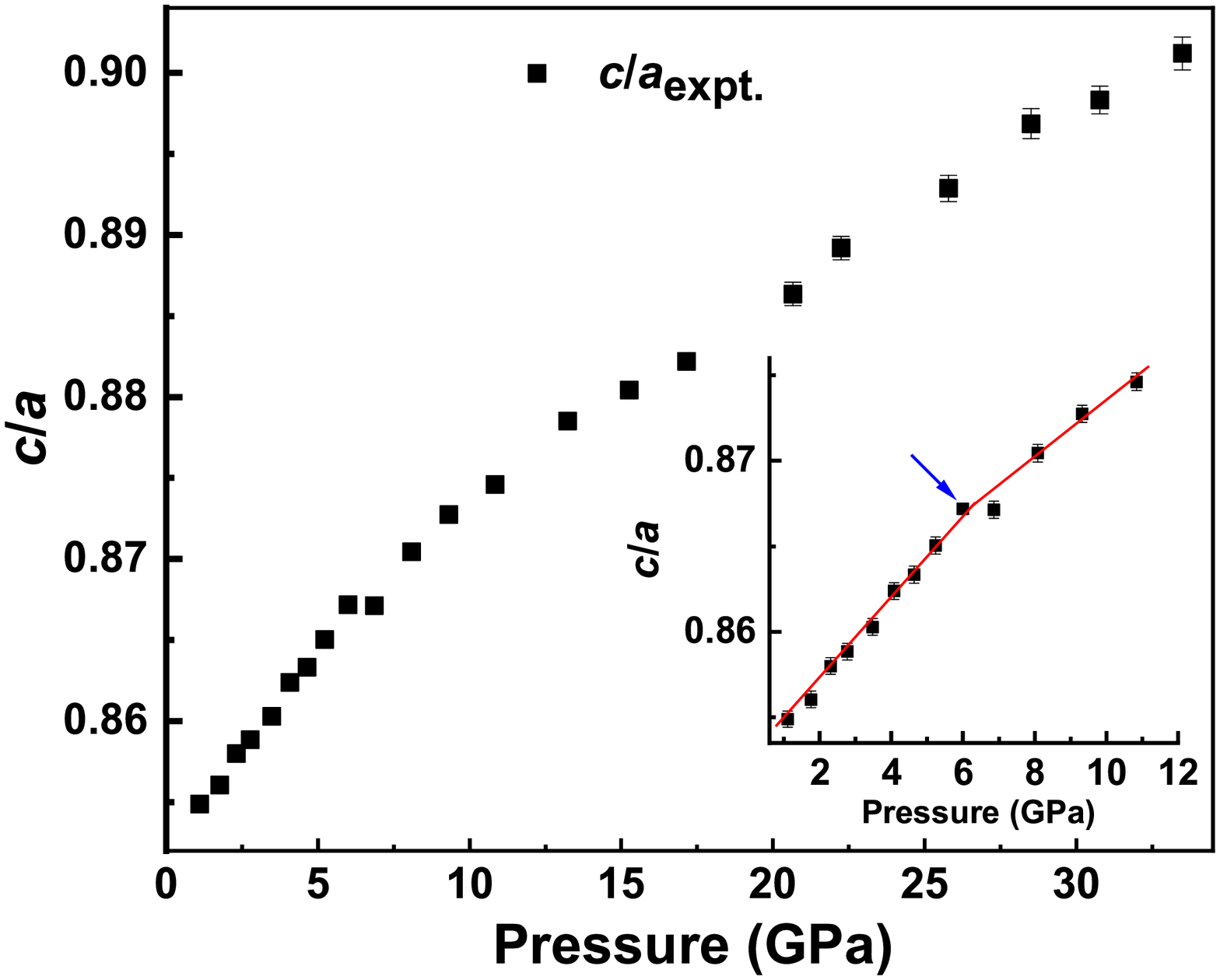}}
\caption{(a) Synchrotron XRD pattern of TlInTe$_2$ at selected pressures. The Bragg peaks, (200) and (211) are indicated to show that their calculated intensities do not fit well with the observed intensity in Rietveld refinement  (b) Pressure versus lattice parameters, $a$ and $c$ of TlInTe$_2$ from both HPXRD and first principles calculations (c) EOS of TlInTe$_2$ is obtained from fitting the P-V (HPXRD and first principles calculations) data to 3$^{rd}$ order BM-EOS. The size of the errors is smaller than the size of the symbols. Pressure versus $c$/$a$ ratio of TlInTe$_2$ from the HPXRD. An obvious slope change of P vs $c$/$a$ ratio at 6 GPa is shown as an inset. The red lines represent guidance to the eyes and a slanted red arrow is to indicate the lattice anomaly during Lifshitz transition.}
\label{fig:XRD}
\end{figure}

To obtain further insight on the nature of transitions observed in high pressure Raman measurements, high pressure synchrotron X-ray diffraction (HPXRD) studies were carried out on TlInTe$_2$. The HPXRD pattern of TlInTe$_2$ for typical pressures (1.1 - 33.5 GPa) as shown in Fig. \ref{fig:XRD}a. The lattice parameters are extracted from the XRD pattern by Le Bail method.\cite{rodriguez1993recent} Our attempts to perform Rietveld refinement were unsuccessful as some of the calculated peak intensities notably (200) and (211) peaks at 5.73$^{\circ}$ and 7.23$^{\circ}$, respectively could not fit well with the experimental pattern [Fig. S2 $\&$ Fig. S3]. Although the refinement of anisotropic displacement parameters of Tl and In has slightly improved the fit, still there is a huge intensity mismatch between the experimental and calculated XRD pattern as shown in Fig. S3. A plausible reason for the intensity discrepancy could be due to a large spatial fluctuation of Tl$^{1+}$ cation inside the Tl$^{1+}$-Te$^{2-}$ Thomson cube resulting in fluctuating electron density at the Tl Wyckoff site, which alters the structure factor of Tl atom. As a result, the intensity and width of calculated XRD pattern do not fit with certain peaks because the effective electron density of X-rays as seen by the Tl$^{1+}$ cation would be different from its actual value owing to the huge spatial fluctuations of Tl$^{1+}$ cation inside the Thomson cube. This argument is further validated by the recent X-ray fluorescence holography (XFH) studies on TlInTe$_2$. The XFH studies reveal three dimensional (3D) atomic images around the Tl atoms are weakly visible at the Tl site whereas the In atoms are completely visible at In site despite the X-ray scattering of Tl (Z=81) is much higher than that of Indium (Z=49).\cite{mimura2011three,hosokawa2015structural} Our HPXRD pattern does not provide any evidence of structural phase transition up to the maximum pressure (33.5 GPa) achieved. Hence, we rule out the possibility of sluggish structural phase transition at 7 GPa proposed by Ves using high pressure Raman spectroscopy.\cite{ves1990high} To further explore the possibility of spatial fluctuations of the Tl$^+$-cation and structural phase transitions, we have carried out CSP calculations using the USPEX package and we have reproduced the B37 phase of TlInTe$_2$ at ambient pressure. The CSP calculations do not predict any structural phase transition up to 33.5 GPa, which is the maximum pressure obtained in HPXRD measurements and the predictions are in good agreement with our present experimental observations (see Fig. S4 and Fig. S5). The computed phonon dispersion curves at ambient as well as at high pressure show dynamical stability of B37 phase (see Fig. S6), which indicates the transitions observed in the Raman measurements are not due to structural phase transitions, which is consistent with our HPXRD measurements. Upon further compression, CSP calculations predict that the B37  phase transforms to cubic (S.G:$Pm\bar{3}m$; Z=4) phase at 50 GPa through an intermediate orthorhombic (S.G:$Pbcm$; Z=4) phase at 37.5 GPa. The 4-fold In-Te$_4$ tetrahedra in B37 phase transforms to a distorted 4-fold In-Te$_4$ tetrahedra with $Pbcm$-type structure, which further transforms to an 8-fold In-Te$_8$ ordered polyhedra in high pressure B2 phase. The $Pbcm$-type structure is a distorted structure of the B2 phase. The structural phase transition sequence thus predicted in TlInTe$_2$ is analogous to that of 1D chain materials such as TlS and TlSe\cite{Demishev1988} except for the symmetry of the distorted intermediate phase [see Fig. S5]. Further detailed discussion on CSP and structural phase transitions are provided in the supporting information (see section S1). In addition, the unit cell parameters and atomic positions of the predicted B37 and high-pressure phases are provided in Table S2 along with the available experimental data.\cite{jana2017intrinsic}

We have also calculated pressure dependent lattice parameters and equation of state (EOS) of TlInTe$_2$. They are plotted together with the experimental data in Fig. \ref{fig:XRD}b $\&$ c for comparison. The calculated static lattice constants are consistent with the HPXRD results at 300 K under high pressure. The pressure dependence of lattice parameter $a$ seems to be steeper than that of $c$, which results from the anisotropic bonding nature of TlInTe$_2$ lattice along the directions of the $a$ and $c$ crystallographic axes [Fig. \ref{fig:XRD}c and Fig. S7]. The anisotropic compression of $a$ and $c$ axes causes an obvious slope change in the pressure versus lattice parameter ratio ($\frac{c}{a}$) at 6 GPa as shown in the inset of Fig. \ref{fig:XRD}d. Despite prominent changes noticed in the pressure dependent phonon modes, the HPXRD studies do not indicate any appreciable change at 19 GPa and 25 GPa. In addition, the experimental EOS shows highly compressible lattice behavior compared to the theoretically predicted EOS. This is especially the case for the pressure range of 3-15 GPa, which is reflected from the low compressible nature of lattice along the $c$-axis from the first principles calculations. Overall there is a good agreement between the theoretical and experimental EOS [Fig. \ref{fig:XRD}c]. To compute the equilibrium bulk modulus (B$_0$), we have fitted both experimental and calculated P-V data to the following 3$^{rd}$ order Birch-Murnaghan equation of state (BM-EOS).\cite{Birch1947}
\begin{equation}
P(V) = \frac{3B_0}{2}\left[\left(\frac{V_0}{V}\right)^{\frac{7}{3}}-\left(\frac{V_0}{V}\right)^{\frac{5}{3}}\right]\Bigg\{1+\frac{3}{4}(B_0'-4)\left[\left(\frac{V_0}{V}\right)^{\frac{2}{3}}-1\right]\Bigg\}
\end{equation}
where P, V$_0$, V, B$_0$ and B$_0$' are pressure, volume at zero pressure, deformed volume, bulk modulus and first derivative of bulk modulus, respectively. The calculated B$_0$ and its first derivative (B$_0'$) determined from the experiment are 19.4$\pm$1.2 GPa and 7.30$\pm$0.29 respectively. The calculated values of B$_0$ (25.55 GPa) and B$_0'$ (5.08) from the theory is overestimated when compared to the experimental values as shown in Fig. \ref{fig:XRD}c. The smaller bulk modulus value suggests a large compressibility of the TlInTe$_2$ lattice resulting from weaker bonding as discussed in the previous section.
\subsection{High pressure transport measurements, electronic structure and Fermi surface topology}
\begin{figure}
\centering
\subfigure[]{\includegraphics[width=3.1in,height=2.8in]{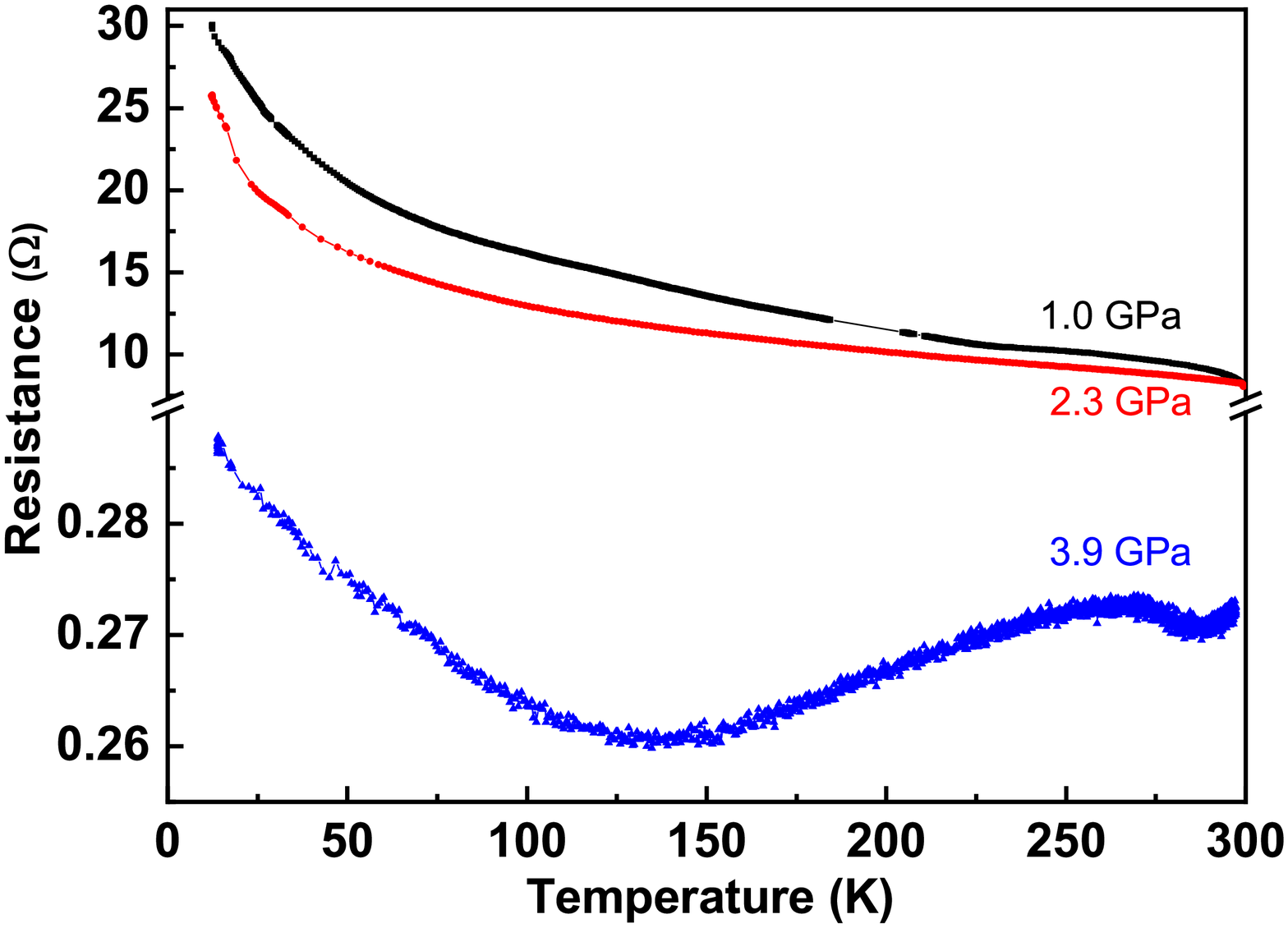}} \hspace{2mm}
\subfigure[]{\includegraphics[width=3.1in,height=2.8in]{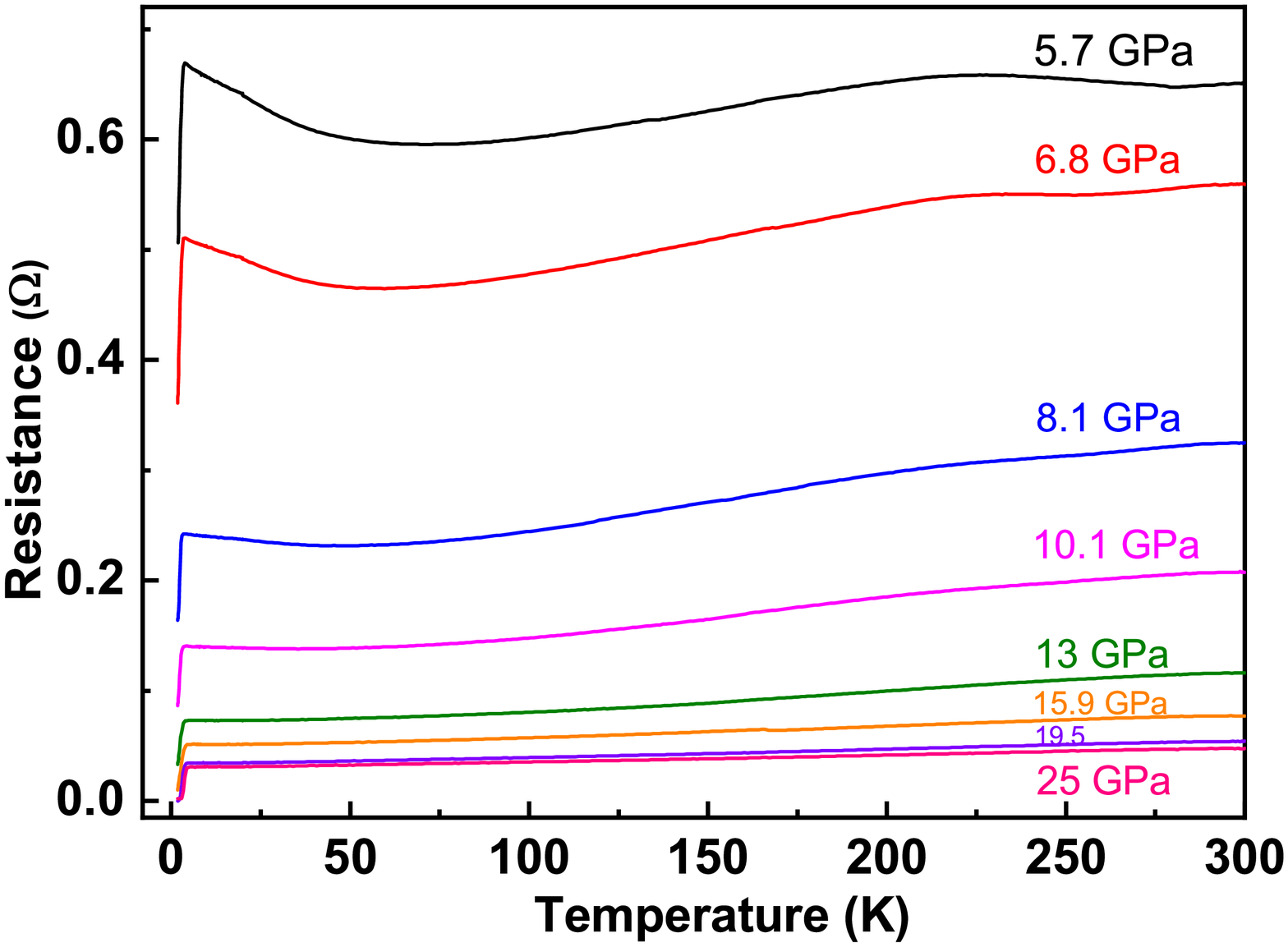}} \vspace{0.1 in}
\subfigure[]{\includegraphics[width=3.1in,height=2.8in]{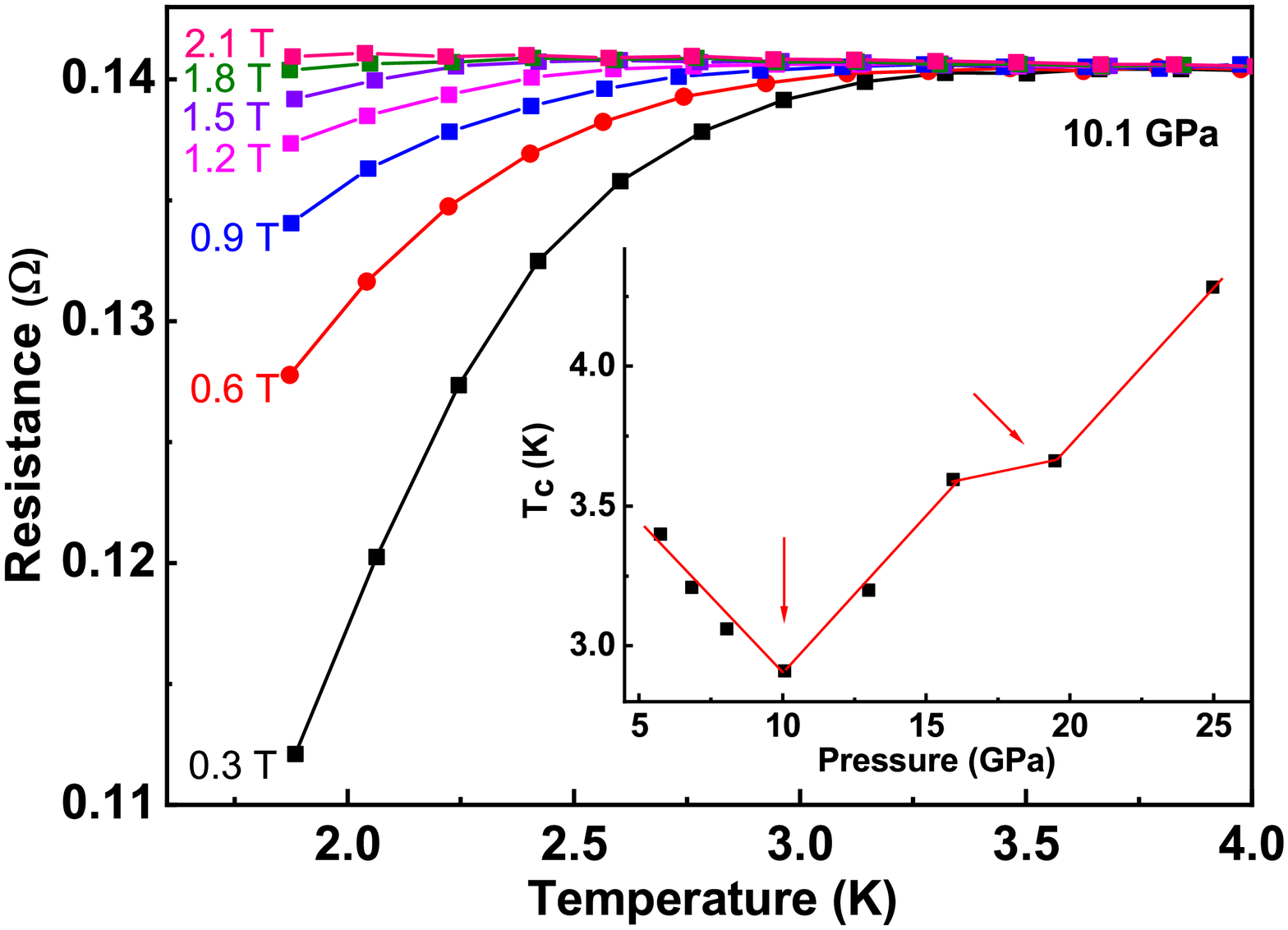}} \hspace{2mm}
\subfigure[]{\includegraphics[width=3.1in,height=2.8in]{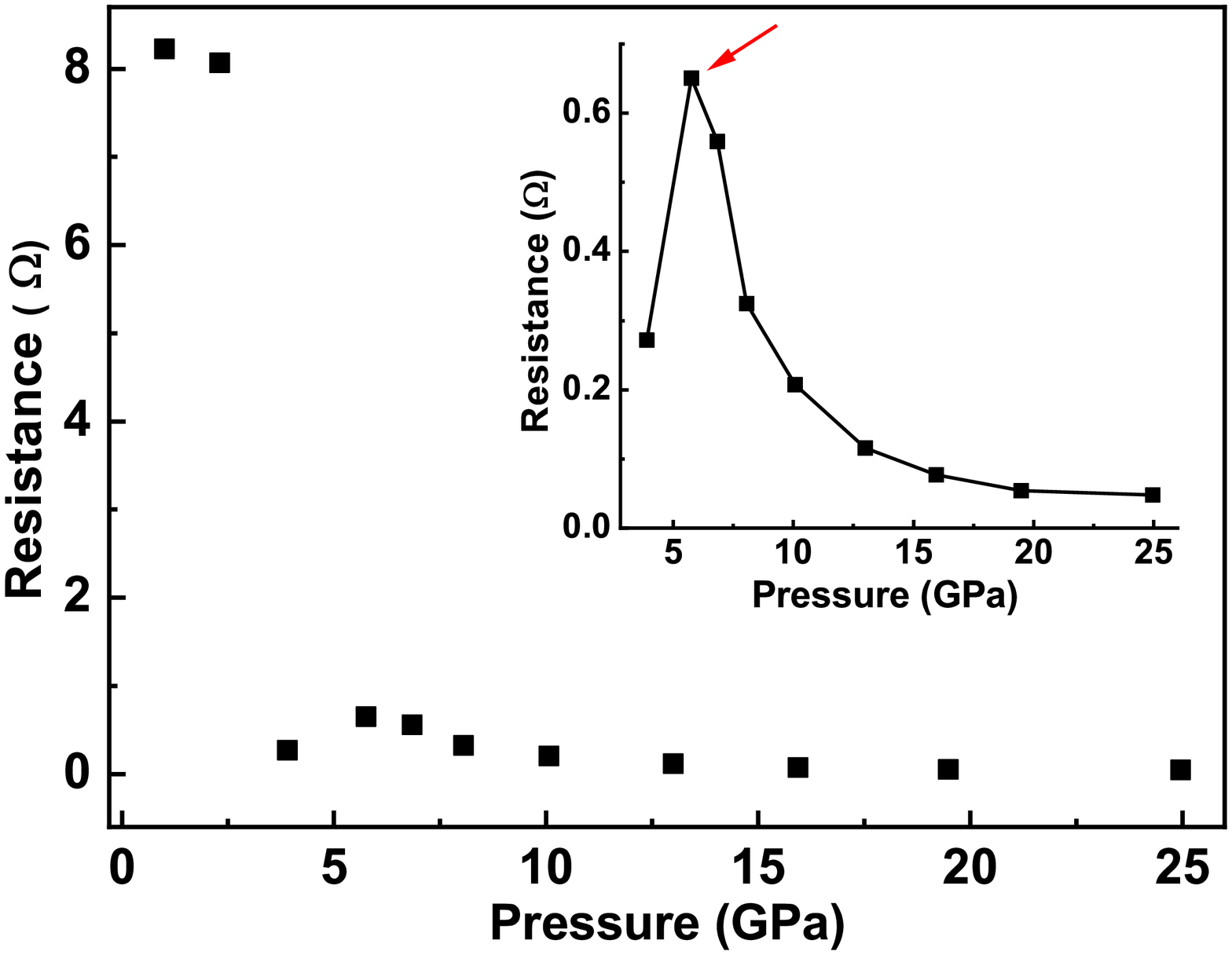}}
\caption{(a) Temperature versus resistance plot of TlInTe$_2$ for low pressures. (b) Temperature versus resistance plot of TlInTe$_2$ for high pressures. (c) Temperature versus resistance for varying magnetic field at 10.1 GPa of TlInTe$_2$ to confirm the superconducting transition. The pressure versus critical superconducting transition temperature (T$_c$) of TlInTe$_2$ is shown as an inset. The red lines represent guidance to the eyes and the slanted red arrows to show the V shaped T$_c$ behavior and the T$_c$ changes at 19 GPa (d) Pressure versus resistance plot of TlInTe$_2$ at room temperature. The pressure versus resistance plot after 3.9 GPa is shown as
an inset. A slanted red arrow is to show the Lifshitz transition at 5.7 GPa}
\label{fig:Transport}
\end{figure}

As high pressure synchrotron XRD studies do not exhibit any obvious signature of structural phase transitions corresponding to the phonon mode softening and frequency collapse at 10, 19 and 25 GPa, transport measurements were carried out to explore further details about the nature of transitions observed in TlInTe$_2$. Figure \ref{fig:Transport}a shows resistance behavior of TlInTe$_2$ determined using a physical property measurement system for selected pressures up to 3.9 GPa and temperatures down to 12 K, and to liquid He temperature for pressures at 5.7 GPa and above. The resistance behavior of TlInTe$_2$ measured at 1.1 GPa evinces a semiconducting behavior [Fig. \ref{fig:Transport}a]. Furthermore, the resistance is suppressed with pressure and a dramatic decrease is observed at 3.9 GPa, which is interpreted as arising from a semiconductor to semimetal transition. Strikingly, the resistance exhibits an increase at 5.7 GPa in comparison to its preceding pressure (3.9 GPa) and the observed resistance anomaly can be related to either topological behavior or Lifshitz transition [Fig. \ref{fig:Transport}d].
A constant resistance behavior is observed around 20 GPa. Upon compression, the resistance drops to near zero at 20 GPa, suggesting a superconducting transition. The smearing out of superconductivity under magnetic field for typical pressures at 5.75 GPa and 10.1 GPa is shown in Fig. \ref{fig:Transport}c and Fig. S8, which confirms the presence of pressure induced superconductivity in TlInTe$_2$. Figure \ref{fig:Transport}c reveals magnetic field dependence of resistance at 10.1 GPa, and the superconductivity disappears completely at 2.1 T. The maximum T$_c$ observed at $\sim$ 25 GPa is 4.3 K. The procedure followed to calculate the T$_c$ value from the resistance plot is shown as an inset of Fig. S8. It is interesting to note the V-shaped T$_c$ behavior of TlInTe$_2$ with the lowest T$_c$ at 10 GPa followed by another distinct anomaly at 19 GPa (inset of Fig. \ref{fig:Transport}c). The experimental plot of upper critical field against temperature at 5.75 GPa and 10.1 GPa can be fitted by using the Ginzburg-Landau (G-L) equation\cite{gao2018,pavlosiuk2015shubnikov}
\begin{equation}
 H_{c2}(T) = H_{c2}(0)\left(1-\frac{T}{T_c}\right)^n
 \end{equation}
where the upper critical field at T = 0 K (H$_{c2}$(0)), and n are fitting parameters. The value of n is found to be $\textless$ 1 and the calculated H$_{c2}$(0) values are 3.6 (at 5.75 GPa) and 4 (at 10.1 GPa). These values are lesser than that of a Bardeen-Cooper-Schrieffer (BCS) weak coupling superconductor with Pauli paramagnetic limit; $\mu_0$ = 1.86 and T$_c$ = 6.34 K at 5.75 GPa.

\begin{figure}
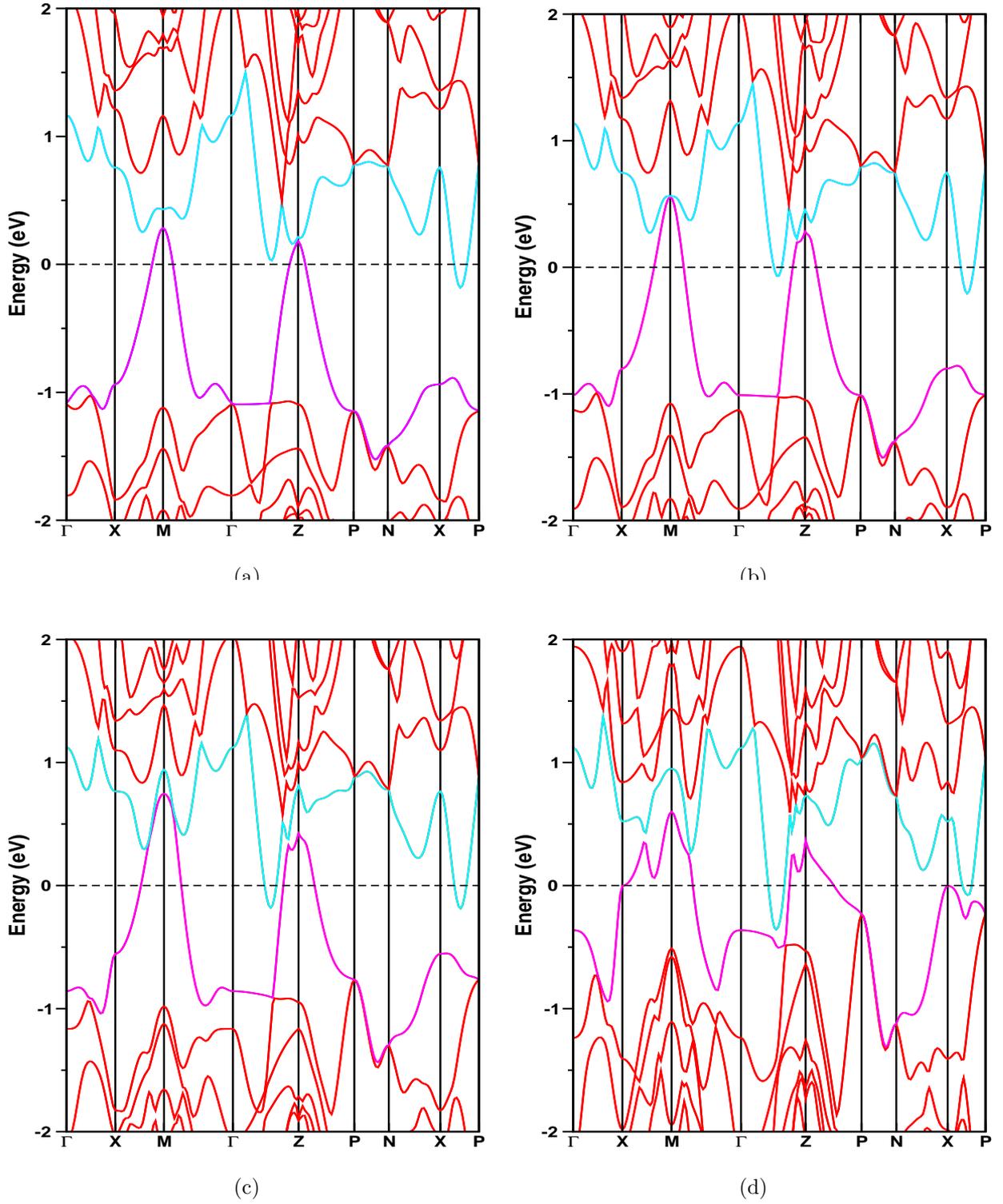

\centering
\subfigure[]{\includegraphics[width=3.1in,height=3.5in]{Figures/Fig4a.eps}} \hspace{2mm}
\subfigure[]{\includegraphics[width=3.1in,height=3.6in]{Figures/Fig4b.eps}} \vspace{0.1 in}
\subfigure[]{\includegraphics[width=3.1in,height=3.5in]{Figures/Fig4c.eps}} \hspace{2mm}
\subfigure[]{\includegraphics[width=3.1in,height=3.5in]{Figures/Fig4d.eps}}
\caption{Calculated electronic band structure of B37 phase for typical pressures (a) 6 GPa (b) 9 GPa (c) 15 GPa and (d) 30 GPa with SOC.}
\label{fig:BS-SO}
\end{figure}

\begin{figure}
    \centering
    \includegraphics[width=7.0in,height=5.in]{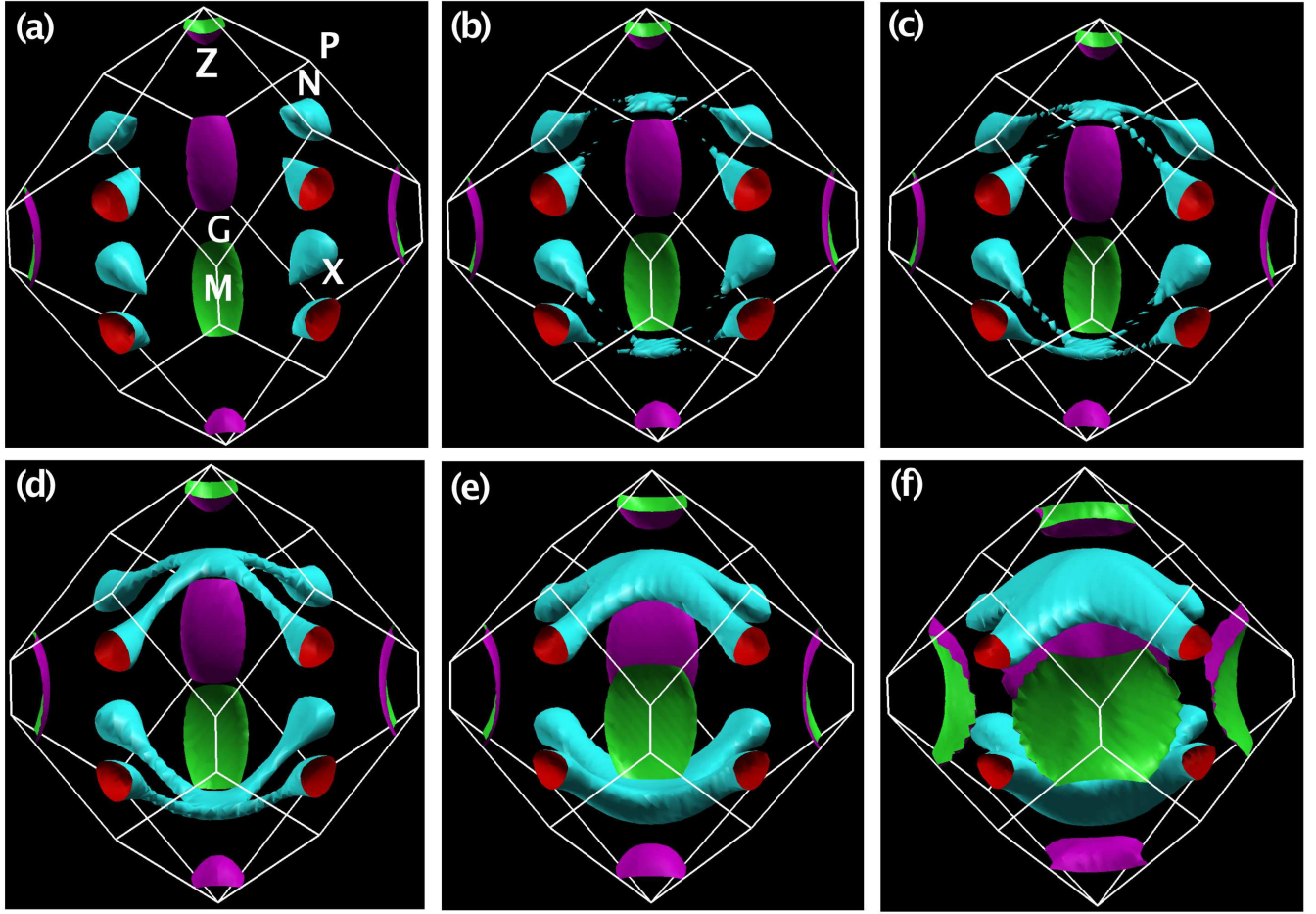}
    \caption{The calculated Fermi surface of B37 phase at different pressures (a) 6.5 GPa (b) 6.8 GPa and (c) 7 GPa and (d) 8 GPa  (e) 15 GPa and (f) 30 GPa with SOC. The new electron pockets are formed above 6.5 GPa and are strengthening with pressure. The new electron pockets are connecting with the other electron pockets in the form of tubular necks to form an umbrella shaped Fermi surface at the top and bottom of the Brillouin zone over the studied pressure range.}
    \label{fig:FS}
\end{figure}

Electronic structure calculations were performed to obtain further insight into the nature of the anomaly noticed at 5.7 GPa (see Fig. \ref{fig:Transport}c), are discussed extensively in this section.  The calculated electronic band structure of TlInTe$_2$ at ambient pressure reveals that it is an indirect band gap semiconductor with the top of valance band maximum (VBM) at M point and the conduction band minimum (CBM) along X-P high symmetry direction. The calculated band gap using TB-mBJ potential at ambient pressure without and with spin-orbit coupling (SOC) are 0.6 eV and 0.35 eV respectively [Fig. S9 and Fig. S10], which are improved over PBE-GGA functional ca. 0.12 eV\cite{jana2017intrinsic} and underestimated when compared to the experimentally observed optical band gap of 0.9-1.1 eV [\cite{panich2008electronic} and refs. therein]. The computed partial density of states (PDOS) of TlInTe$_2$ at ambient pressure is as shown in Fig. S11a. It is inferred from Fig. S11 that the top of the valence band is mainly dominated by Te-$5p$ states and Tl-$6s$ states and they are weakly hybridized with each other. The weak $5p$ electron hybridization of Te atoms and almost isolated $6s$ electronic distribution around Tl atoms impose a moderate constraint on the movement of Tl atoms resulting a strong dynamic abnormality in this material.\cite{Minghui2019}  The minimum of the conduction band is mainly derived from the In-$5s$, Te-$5p$ and Tl-$6p$ states. In general, the states near the Fermi level contribute to the electron-hole transport mechanism from CBM and VBM, respectively. The intrinsic multi-valley degeneracy at Z and along X-P directions at the CBM and along M and Z at VBM is favorable for an efficient electronic transport mechanism and is not intrinsic in commercially available thermoelectric materials, Bi$_2$Te$_3$ and PbTe.\cite{Guangqian2018} Under high pressure, the $5p$ (Te) and $6s$ (Tl)-states from the valence band (at M and Z point of the Brillouin zone) and $5s$-states of In from conduction band (along X-P direction of Brillouin zone) do cross the Fermi level leading to semiconductor $\rightarrow$ semi metal transition around 6 GPa and it becomes metal upon further compression above ca. 17 GPa due to the crossing of more number of states to the Fermi level as illustrated in Figs. \ref{fig:BS-SO}, \ref{fig:FS}, S10 and S11.

The most important discovery in this work is the emergence of superconductivity at 5.7 GPa followed by a V-shaped T$_c$ behavior concomitant with the giant softening of optical phonon mode, A$_g$ at the same pressure, 10-12 GPa [Fig. \ref{fig:Transport}b]. The occurrence of maximum inflection point of resistance noticed at 5.7 GPa is in conjunction with the phonon lifetime change of the A$_g$ mode, the intensity change and compressibility change at 6 GPa. Surprisingly, the changes thus observed are seen to be concurrent with the superconducting transition at 5.7 GPa providing a hint that these transitions can be related to each other. A possibility of topological nontrivial phase has already been predicted for TlSe family. For instance, the first principles calculations on monolayer TlSe reveals an occurrence of topological crystalline insulator state at ambient pressure, which transforms to a topological insulating state under tensile strain\cite{niu2015two}. Similarly, InTe from the same family is predicted to exhibit two successive band inversions at Z and M point of the Brillouin zone at $\sim$1 and $\sim$1.4 GPa respectively\cite{rajaji2018pressure}. A comparison of band structures of TlInTe$_2$ and InTe suggests that the presence of Tl $6s$-states near the Fermi level completely changes the band topology of TlInTe$_2$ when compared to InTe. Moreover, the electronic band structure calculations reveal that the applied pressure simultaneously increases electron and hole populations in the B37 phase as reported for InTe\cite{rajaji2018pressure} and NbAs$_2$\cite{Li2018}. Our calculations do not provide any evidence for the topological transition at 6 GPa, therefore, we anticipate that this transition could be correlated to Lifshitz transition. The increase in FWHM of the A$_g$ phonon mode at 6 GPa might be attributed to the change in topology of the Fermi surface. Changes in the topology of Fermi surface for bulk crystals are explained by Lifshitz in four possible ways.\cite{Lifshitz1960,Kaganov1982} 1) creation or 2) disappearance of a neck in the Fermi surface (the so called Lifshitz neck disruption transition of L2 type); 3) the creation or 4) disappearance of a pocket in the Fermi surface (Lifshitz transition of L1 type). In order to obtain deeper insights into the Lifshitz transitions and the origin of superconductivity, we carried out a detailed analysis of changes in the Fermi surface topology as a function of pressure. As illustrated in Fig. \ref{fig:FS}, the calculated Fermi surface at 6 (Semiconductor to semimetal transition) and 6.5 GPa (see Fig. \ref{fig:FS}a) show an individual hole (magenta) and electron (cyan) pockets. Above 6.5 GPa, the electron pockets are connected through newly emerged spots in the Fermi surface (see figure \ref{fig:FS}b $\&$ c). The connectivity of the electron pockets increases as a function of pressure, along with the appearance of new spots on the Fermi surface. The appearance of new spots and the connecting points of the electron pockets grow as tubular necks, forming an umbrella shaped Fermi surface at the top and bottom of the Brillouin zone, which are ascribed to L1 and L2 type Lifshitz transitions respectively in the pressure range 6.5-8 GPa. Therefore, the combination of L1 and L2 type transitions might be responsible for the emergence of superconductivity in the B37 phase of TlInTe$_2$ as observed in our transport measurements at 5.7 GPa. The neck disruptive L2 type transition induced superconductivity has already been predicted in high temperature H$_3$S superconductor.\cite{Bianconi2015,Jarlborg2016,Durajski2017}

Having resolved the origin of superconductivity, next we focus on the origin of V-shaped T$_c$ behavior at 10 GPa and its correlation to the giant A$_g$ mode frequency softening at 10-12 GPa. HPXRD studies together with structure prediction calculations confirm that the A$_g$ mode softening is unrelated to a structural phase transition. We assign the pressure at which the T$_c$ value goes to a minimum as the critical pressure (P$_c$) [Figure \ref{fig:Transport}]. In order to decipher the frequency softening at 10-12 GPa, we compare our results with iron (Fe) based and PbTaSe$_2$ superconductors. Interestingly, an identical V-shaped T$_c$ behavior has already been extensively studied in Fe-based superconductors such as KFe$_2$As$_2$, RbFe$_2$As$_2$ and CsFe$_2$As$_2$.\cite{tafti2015universal,tafti2014sudden,tafti2013sudden,wang2015upward} The plausible reason for the origin of V-shaped T$_c$ behavior in these materials is interpreted as the change in pairing symmetry from $d$ to $s_\pm$ state before and after the critical pressure (P$_c$), respectively. Similarly, a V-shaped T$_c$ behavior is also reported for PbTaSe$_2$ in the pressure range 0.5-1 GPa, which is attributed to the Lifshitz transition. Moreover, a V-shaped T$_c$ is also predicted for high temperature superconductor H$_3$S in the pressure range 120-500 GPa, which is due to small and unfavorable effect of the applied pressure on electronic density of states and electron-phonon matrix elements.\cite{Durajski2017} In contrast to the iron-based superconductors, TlInTe$_2$ do not possess any magnetic ordering, hence we rule out the possibility of any pairing symmetry change at the critical pressure (P$_c$). Since the giant phonon mode softening and T$_c$ change are observed around the same pressure, we propose that the observed V-shaped T$_c$ behavior is induced by phonons. A phonon mode softening and line width changes similar to that reported here for TlInTe$_2$, has already been reported in cuprate, yttrium barium copper oxide (YbBa$_2$Cu$_3$O$_7$) at low temperature.\cite{stein1996electron,heyen1991two} The phonon mode changes in YbBa$_2$Cu$_3$O$_7$ at low temperature is attributed to the electron-phonon coupling. In order to comprehensively understand the origin of V-shaped T$_c$ behavior in TlInTe$_2$, we carried out a detailed line shape analysis of phonon modes. Surprisingly, an obvious asymmetry begins to emerge at P=8.84 GPa and above to the left (low frequency region) of A$_g$ mode. This asymmetry could not be fitted with a pure Lorentzian, but it could fit very well with a Fano line shape function as shown in Fig. S12 (also see section 4). The fitting of A$_g$ mode with Fano line shape function at 8.84 GPa is shown in Fig. S12. The Fano asymmetry in phonon line shape arises as a result of coherent coupling of Fano interference between discrete phononic states with the broad electronic continuum formed by the conduction electrons. The asymmetric line shape observed in A$_g$ phonon mode is fitted with the Fano line shape using the following equation
\cite{gupta2003laser, cerdeira1973interaction}
\begin{equation}
I(\omega) = A\frac{(q+\epsilon)^2}{1+\epsilon^2}; \hspace{0.3in} where  \hspace{0.15in} \epsilon = \frac{\omega-\omega_0}{\Gamma}
\end{equation}
where A is a constant, $\omega_0$ is the modified  (re-normalized) frequency in the presence of electron-phonon interaction, q is asymmetric parameter, $\frac{1}{q}$ is the strength of the interaction and $\Gamma$(cm$^{-1})$ is the phonon line width, which is related to phonon life time. As $\frac{1}{q}$ tends to zero, the Fano line shape becomes Lorentzian. Fig. S12 represents the fitting of A$_g$ phonon mode with equation (1), yielding q = -10.84(1.00) and $\Gamma$ = 2.10(0.5) cm$^{-1}$. A negative q value represents the interference between discrete phonon states with the electron continuum of the conduction electrons constructively at the lower frequency (cm$^{-1}$) region whereas it destructively interferes at the higher frequency (cm$^{-1}$) region. The large asymmetric parameter (q) values suggests that the electron-phonon interaction strength is weak, however, we anticipate that it might enhance at low temperature close to T$_c$. The pressure dependence of q exhibits a maximum around 10.5 GPa. Similarly, the phonon lifetime (FWHM) of A$_g$ optical mode shows an obvious slope change around 12 GPa. Thus, the giant phonon frequency softening, change in asymmetry parameter and phonon lifetime changes suggest a prominent electron-phonon coupling in TlInTe$_2$ above 8 GPa at room temperature. As illustrated in Fig. S11, the conduction electrons are contributed predominantly by In-$5s$, Tl-$6p$ and Te-$5p$ orbitals at 15 GPa. Since the A$_g$ phonon mode experiences a giant softening at 10-12 GPa, it is most likely the A$_g$ optical phonon mode couples with conduction electrons of In-$5s$ or Te-$5p$ orbitals or both, resulting in V-shaped T$_c$ behavior. Therefore, we propose that the V-shaped T$_c$ behavior in TlInTe$_2$ is induced by electron-phonon coupling.

An abrupt change in slope observed in pressure versus A$_g$ phonon mode at 19 GPa and the T$_c$ change at 20 GPa are very close, so the phonon softening is accompanied with T$_c$ change. The T$_c$ value exhibits a small drop at 20 GPa prior to increasing further at higher pressures. Fig. \ref{fig:Transport}d shows a quadratic decrease in resistance above 5.7 GPa, which saturates around 20 GPa. In order to further understand these changes, we compared our experimental results with the electronic band structure calculations shown in Fig. \ref{fig:BS-SO}. A comparison of Fig. \ref{fig:BS-SO}b $\&$ c suggests that the valence and conduction bands above 15 GPa are overlapping at M point of Brillouin zone whereas they start splitting and begin to overlap with each other along $\Gamma$-Z directions. The overlapping of bands, greater density of states across the Fermi level, and the connecting points growing as tubular connecting necks (see Fig. \ref{fig:FS}) from conduction band, indicate that TlInTe$_2$ undergoes a semimetal to metal transition around 19-20 GPa. Insights into the nature of charge carriers at $\sim$19-20 GPa must await using a more detailed transport studies with single crystals.

Finally, we explore the plausible reasons for the origin of $\sim$4 $\%$ frequency collapse of A$_g$ phonon mode and also a subtle frequency change of E$_g$ mode at $\sim$25 GPa. The HPXRD studies and DFT calculations reveal that the frequency collapse is not accompanied by changes in any of the structural parameters, unit cell volume, lattice parameters and coordination numbers. Since we could not extract bond lengths and bond angles from HPXRD data, we have calculated this information from the first principles calculations. Interestingly, we have noticed that the In-Te bond length becomes stiffens above 25 GPa as shown in Fig. S13. Similarly, a subtle slope change has also been noticed in the bond lengths of In-In and Tl-Tl [Fig. S13].  Hence, the origin of frequency collapse can not be explained by the bond length stiffening occurring at 25 GPa. As shown in Fig. S14, we observed discontinuities in the calculated in-equivalent bond angles (Te-Tl-Te) of distorted Thomson cube of Tl-Te$_8$ at 5-6 GPa and 10-12 GPa. The first discontinuity in the pressure range of 5-6 GPa is due to semi conductor to semi metal transition and the second discontinuity is accompanied with the giant A$_g$ phonon mode softening at 10-12 GPa. Our bond angles calculations do not provide any evidence of drastic change at 25 GPa. So we carried out a detailed structural analysis to ascertain if the frequency collapse is induced by changes in dimensinality or tilting of the Tl-Te$_8$ Thomson cube. As illustrated in Fig. S4, the 1D nature of the ambient B37 phase remains robust even up to 37.5 GPa and a  continuous rotation of square planes of Thompson cubes induces a systematic tilting (see Fig. S15). However, the continuous tilting does not justify an abrupt frequency change (collapse) at 25 GPa. Therefore, we rule out the possibility of either 1D (chains) to 2D (sheets/planes) or tilting of Tl-Te$_8$ changes as driving force for the cause of the observed frequency collapse at 25 GPa. Therefore, we anticipate that the plausible origin of the frequency collapse at 25 GPa might be stemmed from some other unusual phenomena and more detailed studies are required to resolve its origin.
\section{Conclusions}
We systematically investigated the origin of pressure induced superconductivity, giant softening and frequency collapse of A$_g$ mode using high pressure Raman spectroscopy, synchrotron XRD, transport measurements, crystal structure prediction and first principles calculations. A semiconductor to semimetal transition is observed around 4 GPa followed by a resistance anomaly at 6 GPa, which is correlated to the Lifshitz transition. A superconducting transition emerged at 5.7 GPa with a T$_c$ of 3.8 K and the appearance of additional electron pockets in the Fermi surface at 6.8 GPa confirms the superconductivity is induced by  L1 and L2-type Lifshitz transitions owing to the changes in the topology of the Fermi surface. A V-shaped T$_c$ behavior accompanied with a giant A$_g$ phonon mode softening at 10-12 GPa is ascribed to the interaction between the discrete phonon states with electronic continuum of conduction electrons from the In-5s or Te-5p (or both) conduction electrons causing Fano line shape in A$_g$ mode due to electron-phonon coupling. A semimetal to metal transition at 19-20 GPa is seen to be accompanied with an obvious T$_c$ change. Finally, a plausible origin of $\sim$4 $\%$ frequency collapse of A$_g$ phonon mode coupled with In-Te and Tl-Tl/In-In bond length changes at 25 GPa remains an unresolved puzzle. Our study provides a new direction to unveil unprecedented pressure induced novel phenomena in TlSe family. The knowledge of exotic quantum phases observed at moderate pressures can be used to synthesize materials with outstanding thermoelectric and quantum behaviors at ambient conditions by chemical pressure.

\section{Acknowledgments}
Y. D acknowledges the support from National Key Research and Development Program of China 2018YFA0305703; The National Natural Science Foundation of China (NSFC)-U1930401, 11874075, and Science Challenge Project TZ2016001. High pressure synchrotron XRD studies were performed at HPCAT (Sector 16), Advanced Photon Source (APS), Argonne National Laboratory. HPCAT operations are supported by DOE-NNSA’s Office of Experimental Sciences. The APS is a U.S. Department of Energy (DOE) Office of Science User Facility operated for the DOE Office of Science by Argonne National Laboratory under Contract No. DE-AC02-06CH11357. We acknowledge Sergey Tkachev, GSECARS, APS, for Ne gas loading in the DAC. NYK would like to thank the Science and Engineering Research Board and Indo-US Scientific Technology Forum and National Science Foundation (EAR-1723160) for funding. NYK would like to acknowledge Prof. Artem R. Oganov (supervisor) and Dr. M. Mahdi Davari Esfahani for their support and discussions and also Stony Brook University for providing computational resources. YAS acknowledges HPSTAR for post-doctoral fellowship. KB acknowledges Sheikh Saqr Career Fellowship and Department of Science $\&$ Technology, India for partial support. \\
$^*$\emph{Author for Correspondence, E-mail: \\
Y. A. Sorb: sorubya@gmail.com \\
N. Yedukondalu: nykondalu@gmail.com \\
Yang Ding: yang.ding@hpstar.ac.cn}

\bibliographystyle{plain}
\bibliography{references}

\clearpage

{}

\end{document}


\begin{center}
{\Huge{\bf Supporting Information}}
\end{center}

\title{Origin of superconductivity and giant phonon softening in TlInTe$_2$ under pressure}
\author{Sorb Yesudhas$^{1*}$, N. Yedukondalu$^{2*,3}$, Manoj K. Jana$^4$, Jianbo Zhang$^1$, Jie Huang$^1$, Bijuan Chen$^1$, Hongshang Deng$^1$, Raimundas Sereika$^{1,5}$, Hong Xiao$^1$, Stanislav Sinogeikin$^6$, Curtis Kenney-Benson$^7$, Kanishka Biswas$^4$, John B. Parise$^{2,8}$, Yang Ding$^{1*}$, Ho-kwang Mao$^1$}
\affiliation{$^1$Center for High-Pressure Science $\&$ Technology Advanced Research, Beijing 100094, P.R. China
 $^2$Department of Geosciences, Center for Materials by Design, and Institute for Advanced Computational Science, State University of New York, Stony Brook, New York 11794-2100, USA
 $^3$Rajiv Gandhi University of Knowledge Technologies, Basar, Telangana-504107, India
 $^4$New Chemistry Unit, Jawaharlal Nehru Centre for Advanced Scientific Research (JNCASR), Bangalore, India
 $^5$Vytautas Magnus University, K. Donelai\v{c}io Street 58, Kaunas 44248, Lithuania
 $^6$DAC Tools, Custom Equipment for High-Pressure Research, Naperville, IL 60565-2925, USA
 $^7$HPCAT, X-ray Science Division, Advanced Photon Source, Argonne National Laboratory, Lemont IL, 60439, USA
 $^8$National Synchrotron Light Source II, Brookhaven National Laboratory, Upton, New York 11973, USA
 }

\maketitle

\begin{table}
\caption{The experimental Raman mode frequencies, pressure coefficients and Gruneisen parameters of A$_g$ and E$_g$ modes of TlInTe$_2$ are calculated by fitting pressure versus Raman frequencies to the equation $\omega(P)$ = $\omega(P_0$) + a $\times$(\ P-P$_0$), where {$\it P_0$} is the ambient pressure and 'a' is the pressure coefficient.}
\begin{ruledtabular}
\begin{tabular}{cccc}
Symmetry type    &   $\omega(P_0)$  $(cm^{-1})$    &   a $(cm^{-1} GPa ^{-1})$   &     $\gamma$     \\ \hline
A$_g$           &       127.5             &    1.51                      &       0.23      \\
E$_g$            &       139.1             &    3.54                     &       0.5       \\

\end{tabular}
\end{ruledtabular}
\label{tab:table1}
\end{table}

\begin{table}[tbp]
\caption{Crystal structure prediction of ambient and high pressure phases of TlInTe$_2$ and are compared with the available experimental data.\cite{jana2017intrinsic}}
\label{table2}
\begin{ruledtabular}
\begin{tabular}{ccccccc}
 Phase            &   Pressure   & Lattice constants  &     Atom (Wyckoff site)     &     x     &   y    &    z  \\
                  &   (GPa)      &  (\AA)        &                        &           &        &        \\ \hline
 $I4/mcm$ (B37)   &    0         &  a = 8.408    &     Tl ($4a$)          &  0.0000    &   0.0000  &   0.2500  \\
  (Z=4)           &              &  c = 7.174    &     In ($4b$)          &  0.0000    &   0.5000  &   0.2500  \\
                  &              &               &     Te ($8h$)          &  0.1855  &  0.6855  &   0.0000  \\ \\

$I4/mcm$ (B37)\cite{jana2017intrinsic}    &    0    &  a = 8.47980   &     Tl ($4a$)          &  0.0000    &   0.0000  &   0.2500  \\
(Z=4)             &              &  b = 7.18894    &     In ($4b$)          &  0.0000    &   0.5000  &   0.2500   \\
                  &              &                 &     Te ($8h$)          &  0.18098  &  0.68098  &   0.0000  \\   \\

$Pbcm$            &    40        &  a = 3.543    &     Tl ($4d$)          &  0.0141    &   0.3100  &   0.2500  \\
 (Z=4)            &              &  b = 19.185    &     In ($4d$)          &  0.0038    &   0.0661  &   0.2500  \\
                  &              &  c = 4.789    &     Te1 ($4d$)          &  0.4142    &  0.4442   &   0.2500  \\
                  &              &               &     Te2 ($4d$)          &  0.5336    &  0.1821   &   0.2500  \\                   \\

$Pm\bar{3}m$      &    50        &  a = 8.408    &     Tl1 ($1b$)          &  0.5000    &   0.5000  &   0.5000  \\
(Z=4)             &              &               &     Tl2 ($3c$)          &  0.5000    &   0.0000  &   0.5000  \\
                  &              &               &     In1 ($1a$)          &  0.0000    &   0.0000  &   0.0000  \\
                  &              &               &     In2 ($3d$)          &  0.0000    &   0.5000  &   0.0000  \\
                  &              &               &     Te ($8h$)          &  0.2457     &  0.7543  &  0.7543  \\
\end{tabular}
\end{ruledtabular}
\newline
\label{tab:table2}
\end{table}

\clearpage

\begin{figure}
\centering
\subfigure[]{\includegraphics[width=3.1in,height=2.5in]{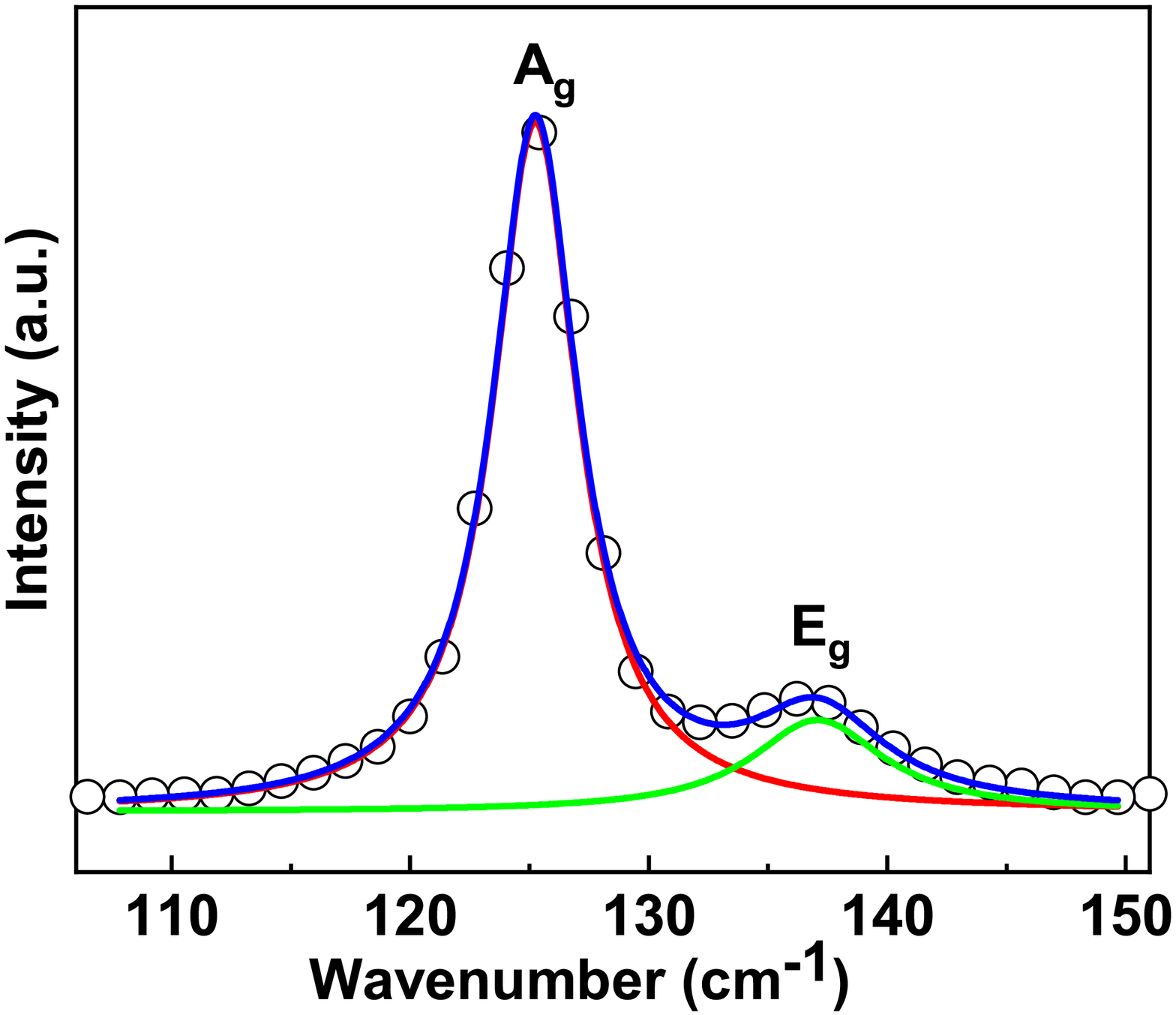}} \hspace{2mm}
\subfigure[]{\includegraphics[width=3.1in,height=2.5in]{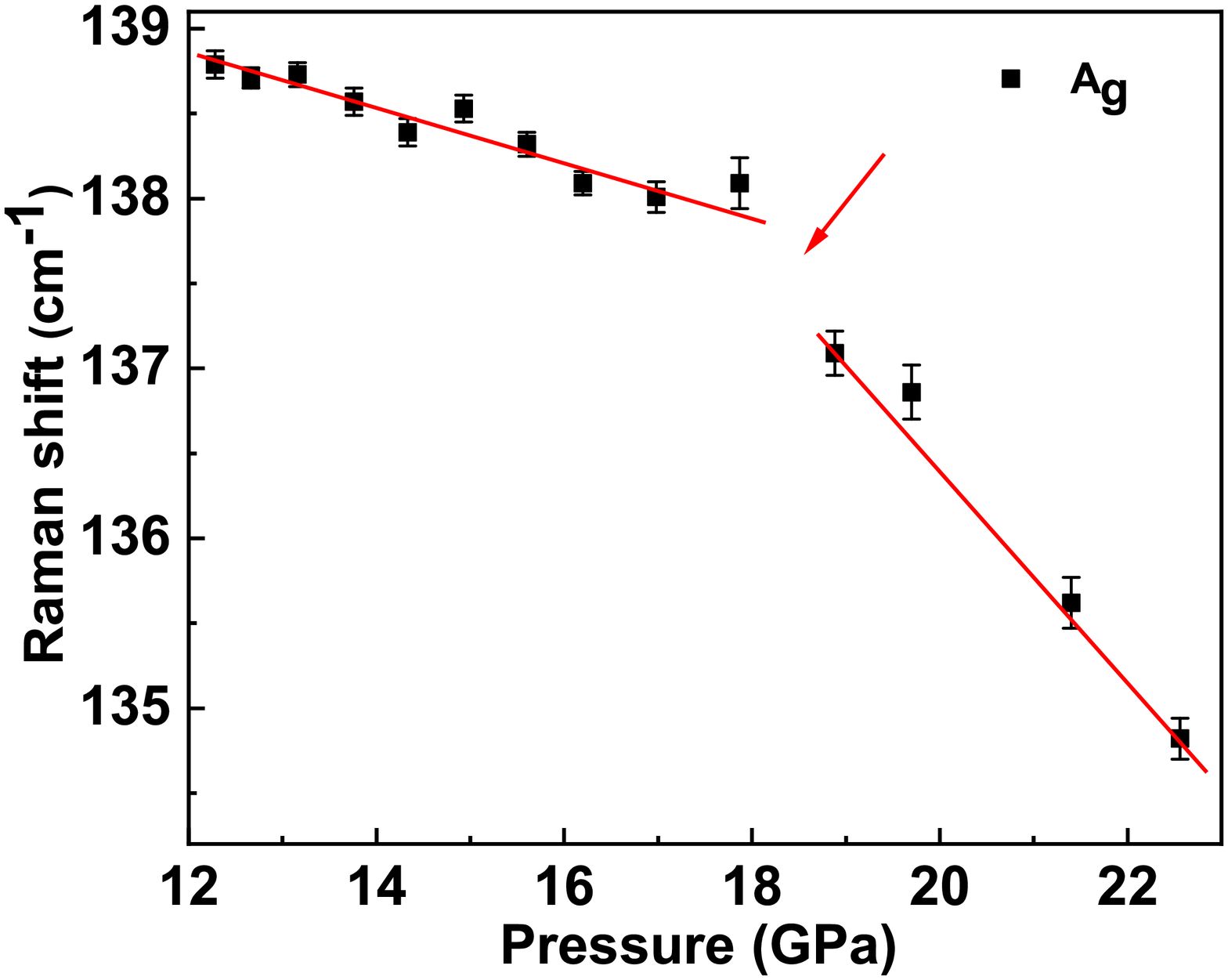}} \vspace{0.1 in}
\caption{(a) Raman spectrum of TlInTe$_2$ at ambient pressure. (b) Pressure vs frequency of A$_g$ mode in the pressure range of 12-23 GPa. The red arrow indicates phonon discontinuity at around 19 GPa.}
\label{fig:XRDfit}
\end{figure}
\begin{figure}
\centering
\subfigure[]{\includegraphics[width=3.1in,height=2.5in]{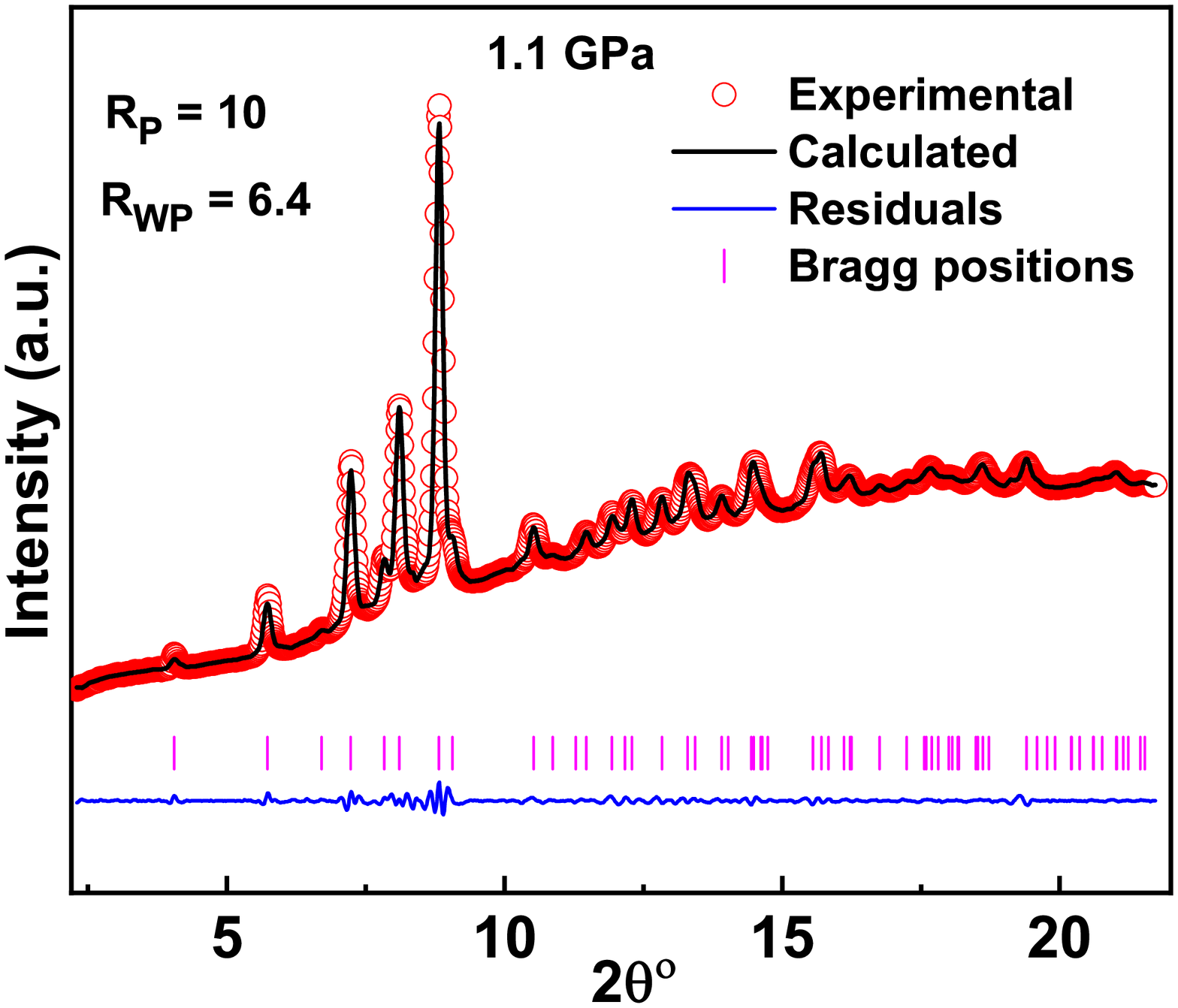}} \hspace{2mm}
\subfigure[]{\includegraphics[width=3.1in,height=2.5in]{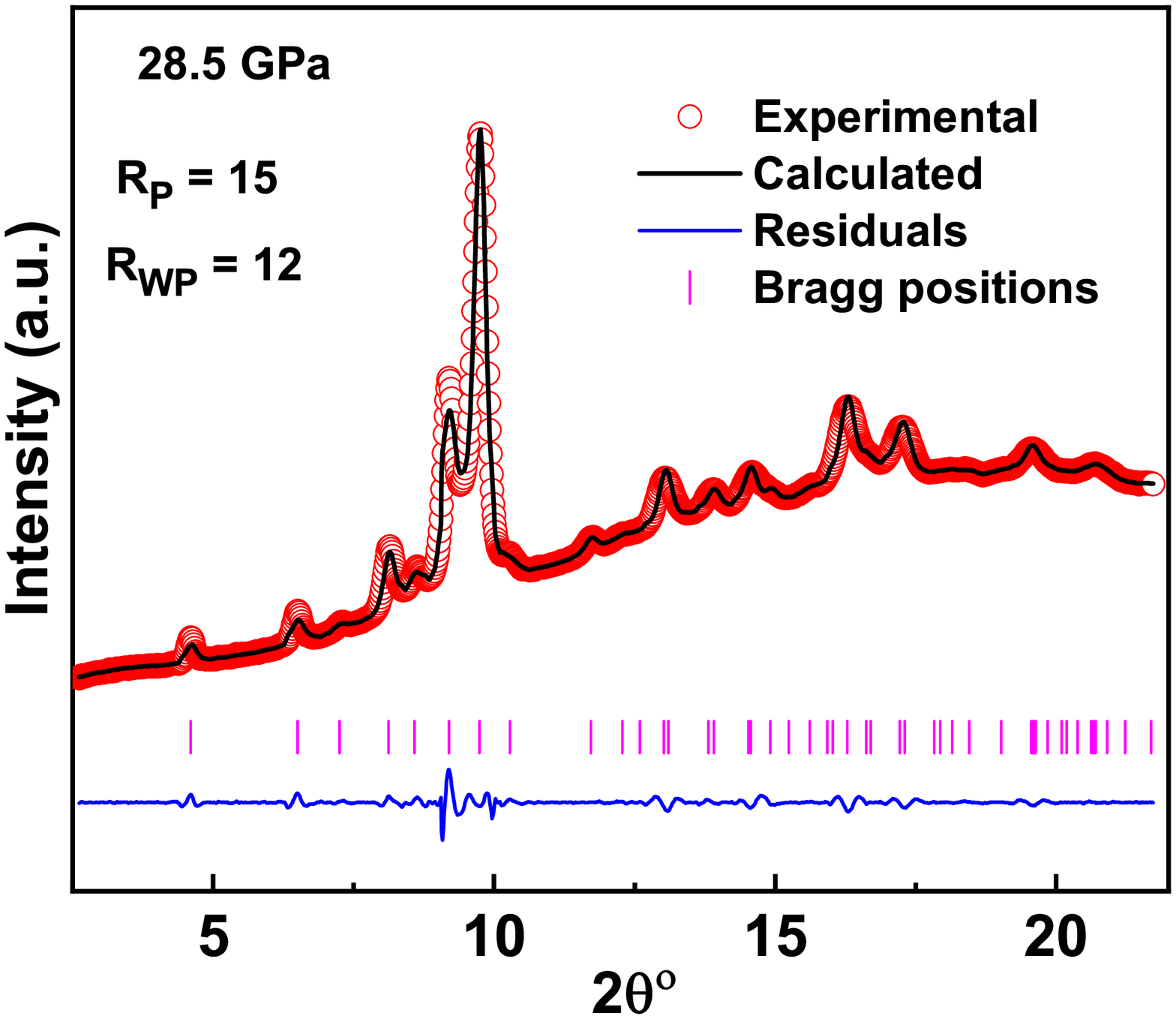}} \vspace{0.1 in}
\caption{Le Bail fitting of synchrotron XRD pattern of TlInTe$_2$ at (a) 1.1 GPa (b) 28.5 GPa.}
\label{fig:XRDfit}
\end{figure}

\begin{figure}
\centering
\subfigure[]{\includegraphics[width=3.1in,height=2.5in]{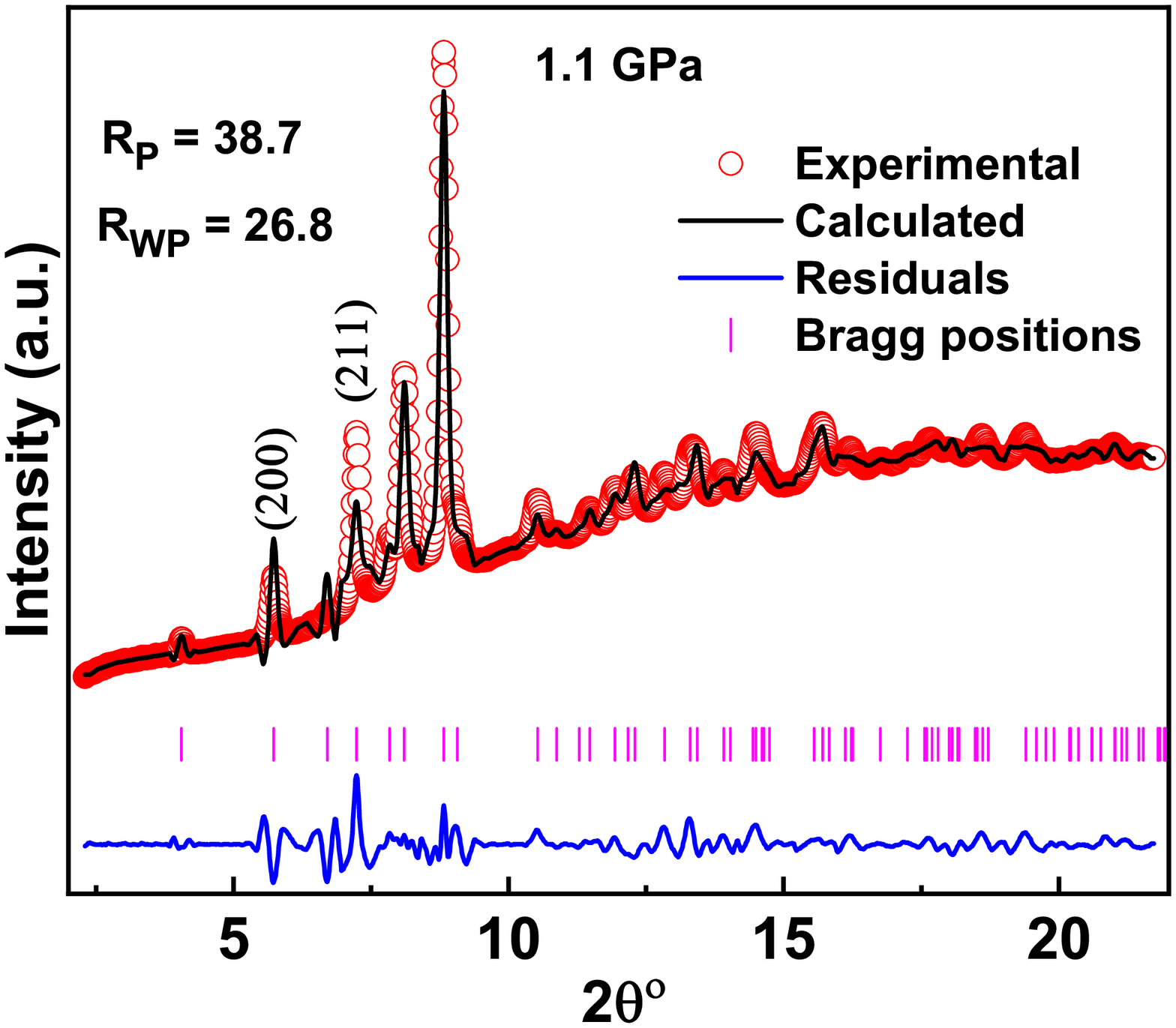}} \vspace{0.1 in}
\caption{Rietveld refinement of synchrotron XRD pattern of TlInTe$_2$ at 1.1 GPa. The intensity of (200), (211) and some of the higher angle peaks do not fit with the experimental pattern.}
\label{fig:XRDfit}
\end{figure}
\begin{figure}
    \centering
    \includegraphics[width=5.0in,height=3.5in]{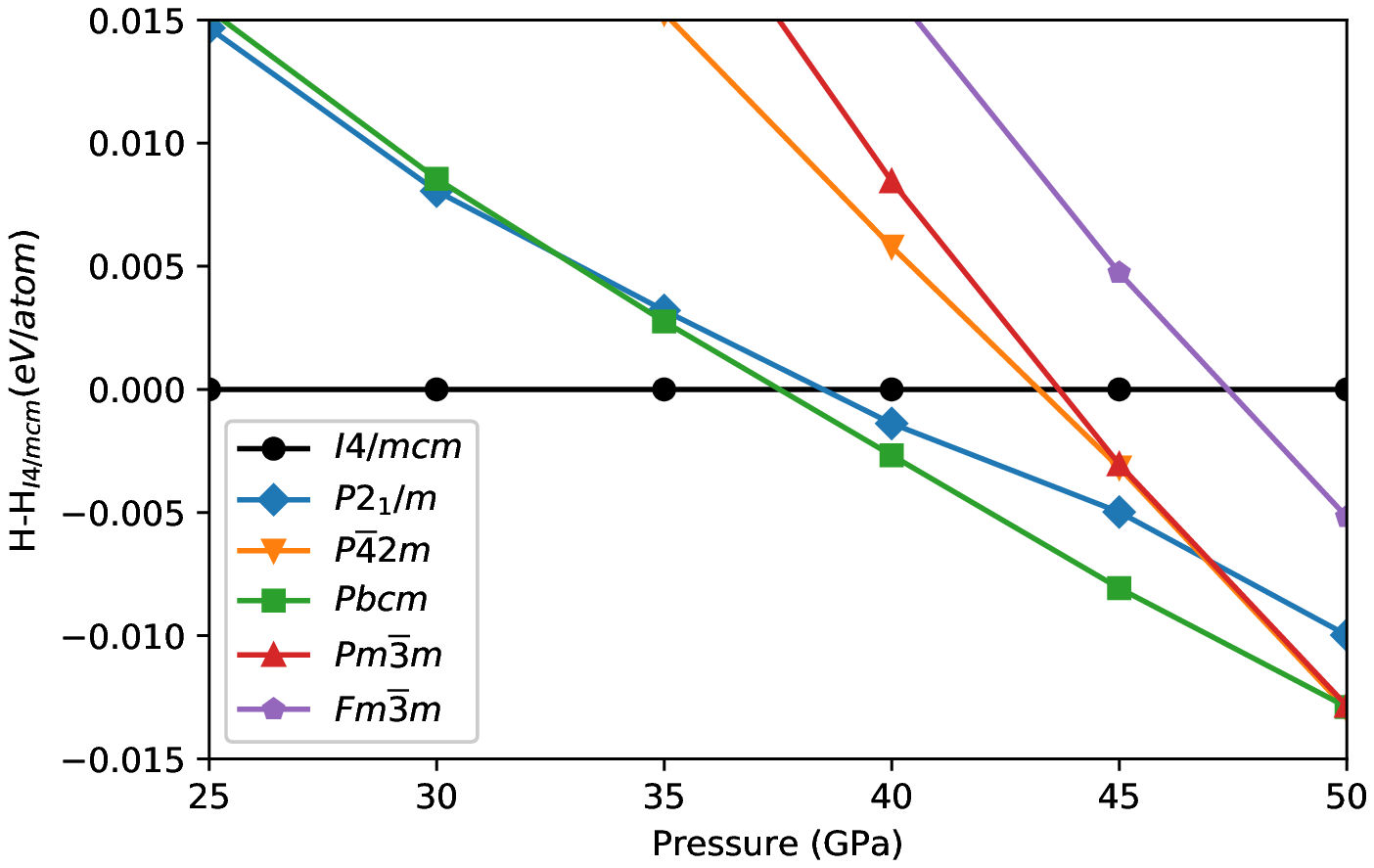}
    \caption{Calculated enthalpy difference of the predicted high pressure phases relative to the ambient B37 phase as a function of pressure. The predicted structural transition sequence in TlInTe$_2$ as follows: $I4/mcm$ $\rightarrow$ $Pbcm$ $\rightarrow$ $Pm\bar{3}m$ in the pressure range 0-50 GPa.}
    \label{fig:enthalpy}
\end{figure}

\begin{figure}
    \centering
    \includegraphics[width=7.0in,height=3.5in]{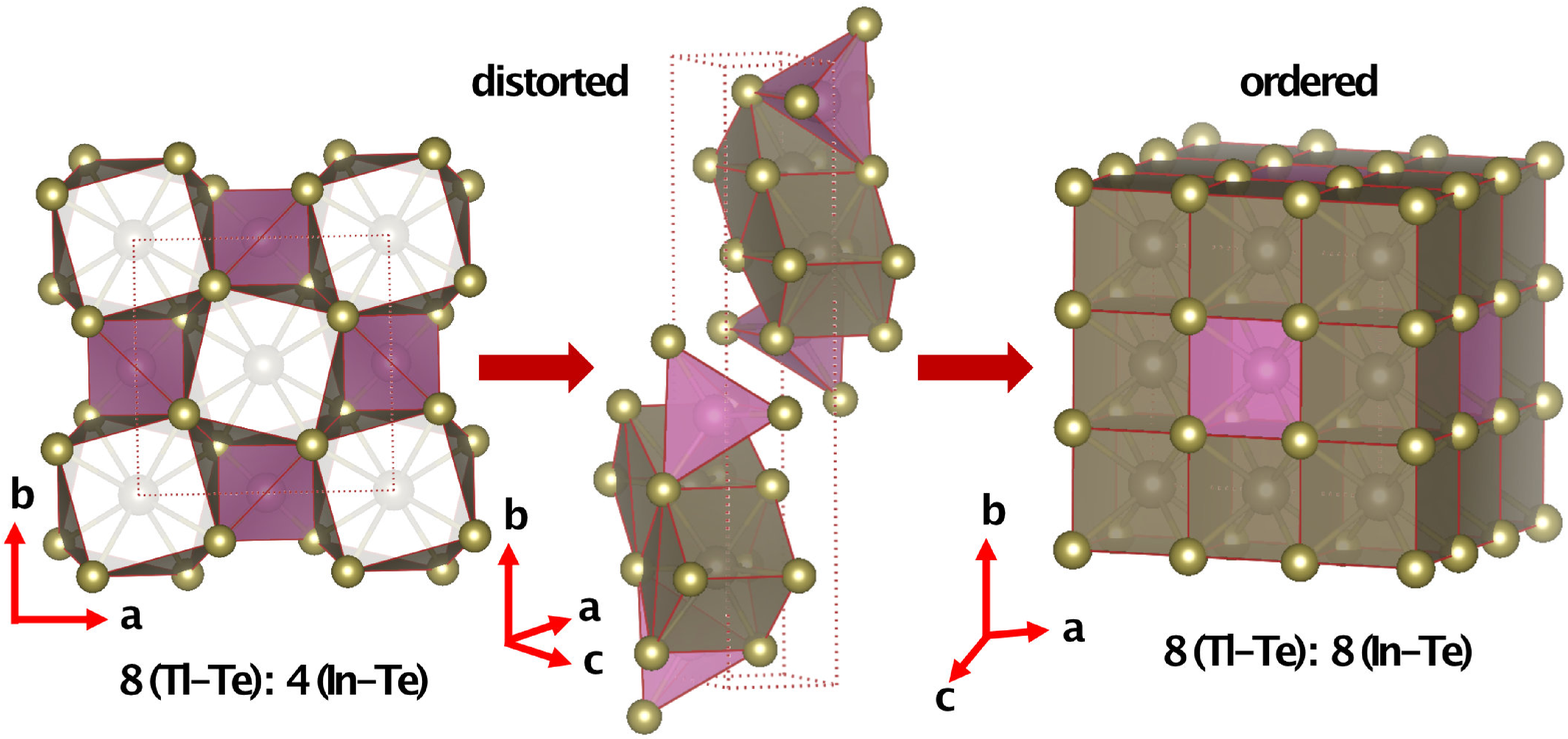}
    \caption{Crystal structures of predicted ambient $I4/mcm$ (B37) (left), intermediate $Pbcm$-type (middle) and high pressure $Pm\bar{3}m$ (B2) (right) phases of TlInTe$_2$.}
    \label{fig:str}
\end{figure}
\clearpage
\begin{figure}
    \centering
    \includegraphics[width=5.0in,height=3.5in]{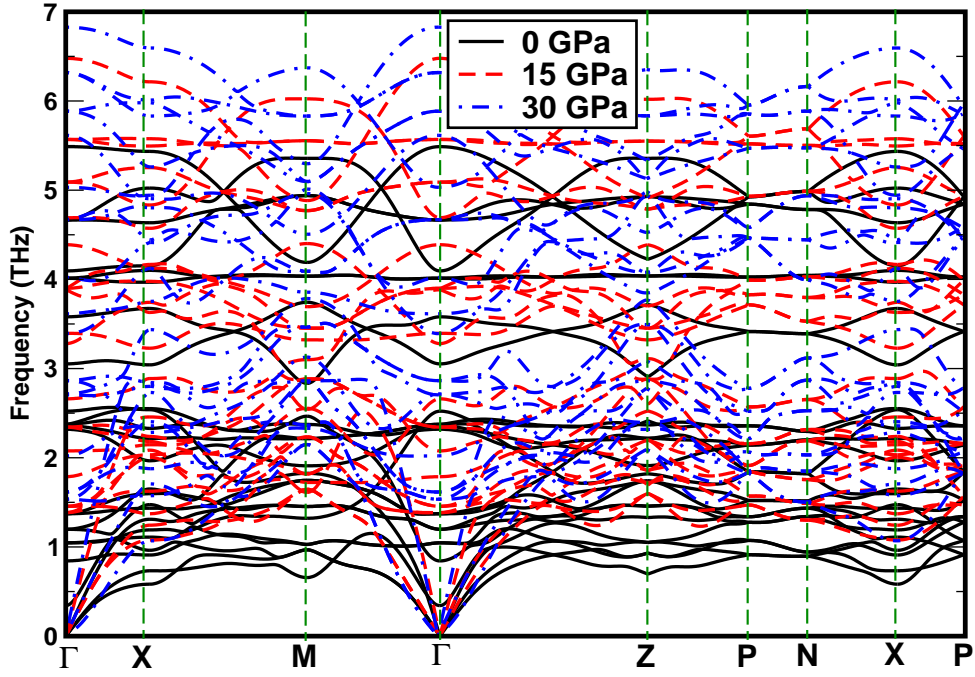}
    \caption{Calculated phonon dispersion curves as a function of pressure for B37 phase, which shows the dynamical stability of B37 at 0, 15 and 30 GPa pressures.}
    \label{fig:PD}
\end{figure}

\begin{figure}
    \centering
    \includegraphics[width=5.0in,height=3.5in]{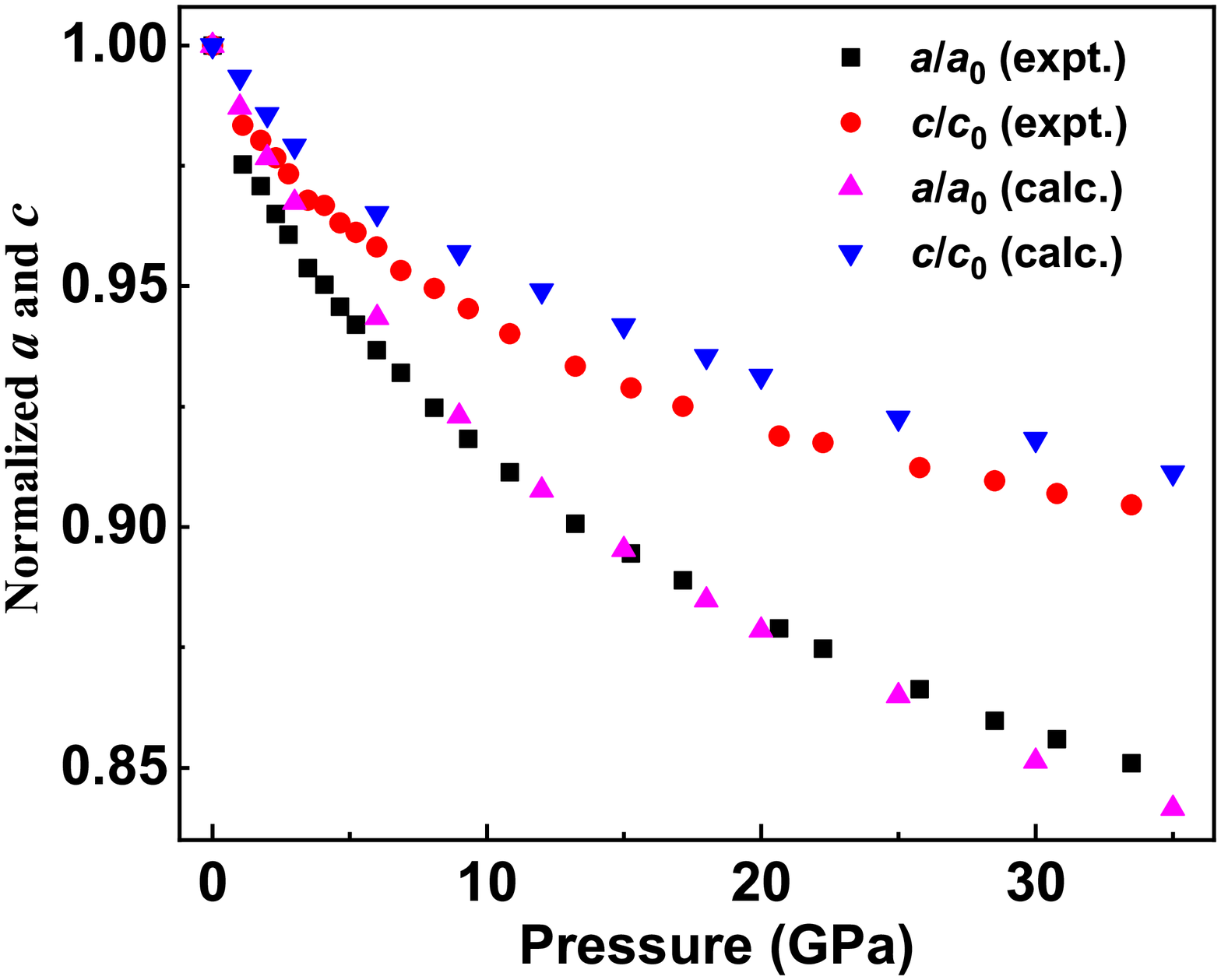}
    \caption{Pressure vs normalized lattice constants, $a$ and $c$ of B37 phase of TlInTe$_2$ from both experiment and theory.}
    \label{fig:Nac}
\end{figure}

\clearpage
\section{Crystal structure prediction at high pressure}
In order to compliment the experimental observations, first principles crystal structure predictions have been performed using USPEX package at different pressures $\sim$0, 10, 30 and 50 GPa with 2, 4 and 8 f.u. per unit cell. Our prediction of crystal structure of the ambient phase is consistent with the experimental observation and also with the previous XRD results.\cite{jana2017intrinsic} The ambient tetragonal structure with space group $I4/mcm$ (B37) is found to be stable up to 37.5 GPa, which is consistent with the present HPXRD measurements. The B37 phase undergoes a pressure induced structural phase transition to a body centred cubic $Pm\bar{3}m$ (B2) phase at 50 GPa through an intermediate distorted orthorhombic ($Pbcm$) phase at 37.5 GPa. The structural transition sequence predicted in this work is closely comparable to the structural transition sequences (B37 $\rightarrow$ distorted $\alpha-NaFe_2O$-type $\rightarrow$ B2) of TlS and TlSe,\cite{Demishev1988} except for the symmetry of an intermediate phase. Interestingly, the presence of Tl$^+$ atom in TlInTe$_2$ provides greater stability for B37 phase over wide pressure range (0-37.5 GPa) when compared to that of iso-structural InTe. InTe undergoes a series of structural phase transitions (B37 $\rightarrow$ B1 $\rightarrow$ B2) below 15 GPa.\cite{rajaji2018pressure} The B37 (Z=8) phase of InTe transforms to a 6-fold coordinated B1 phase (Z = 1) at $\sim$ 6.5 GPa and with further compression, it transforms to 8-fold coordinated B2 phase (Z = 4). In contrast to this, the B37 (Z = 4) phase of TlInTe$_2$ transforms to a distorted $Pbcm$-type structure, with further compression, it transforms to B2 (Z = 4) phase and might be to B1 (Z = 4) phase under very high pressure. Moreover, we also predicted meta stable phases such as $P2_1/m$, $P\bar{4}2m$ and $Fm\bar{3}m$ (B1) at T = 0 K and they might be energetically competitive, if the kinetic effects are considered under high pressure. Thus the predicted crystal structures provide very useful information on crystal structure of TlSe family for future experimental and theoretical studies at high pressure.

\begin{figure}
\centering
\subfigure[]{\includegraphics[width=3.1in,height=2.5in]{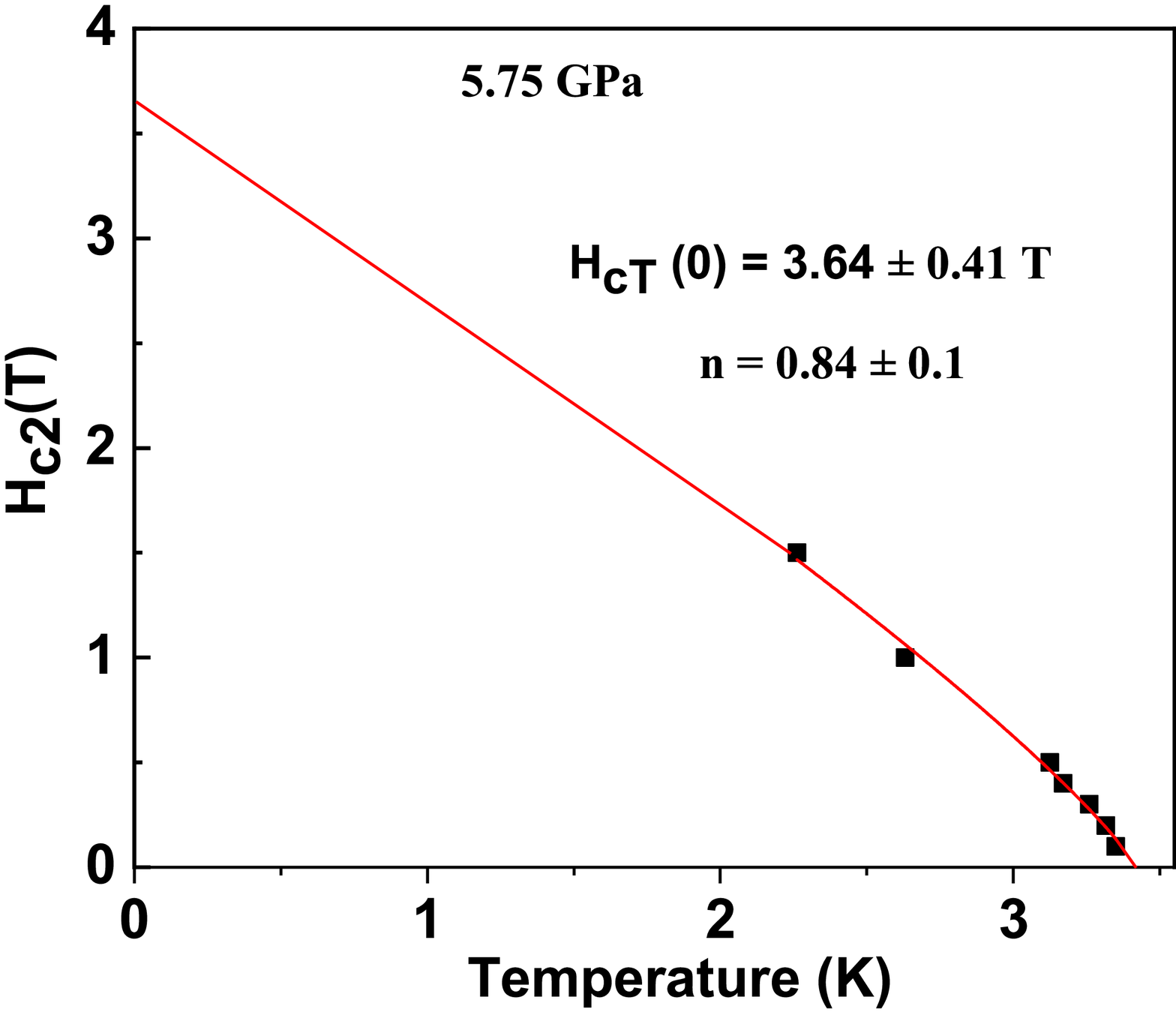}} \hspace{2mm}
\subfigure[]{\includegraphics[width=3.1in,height=2.5in]{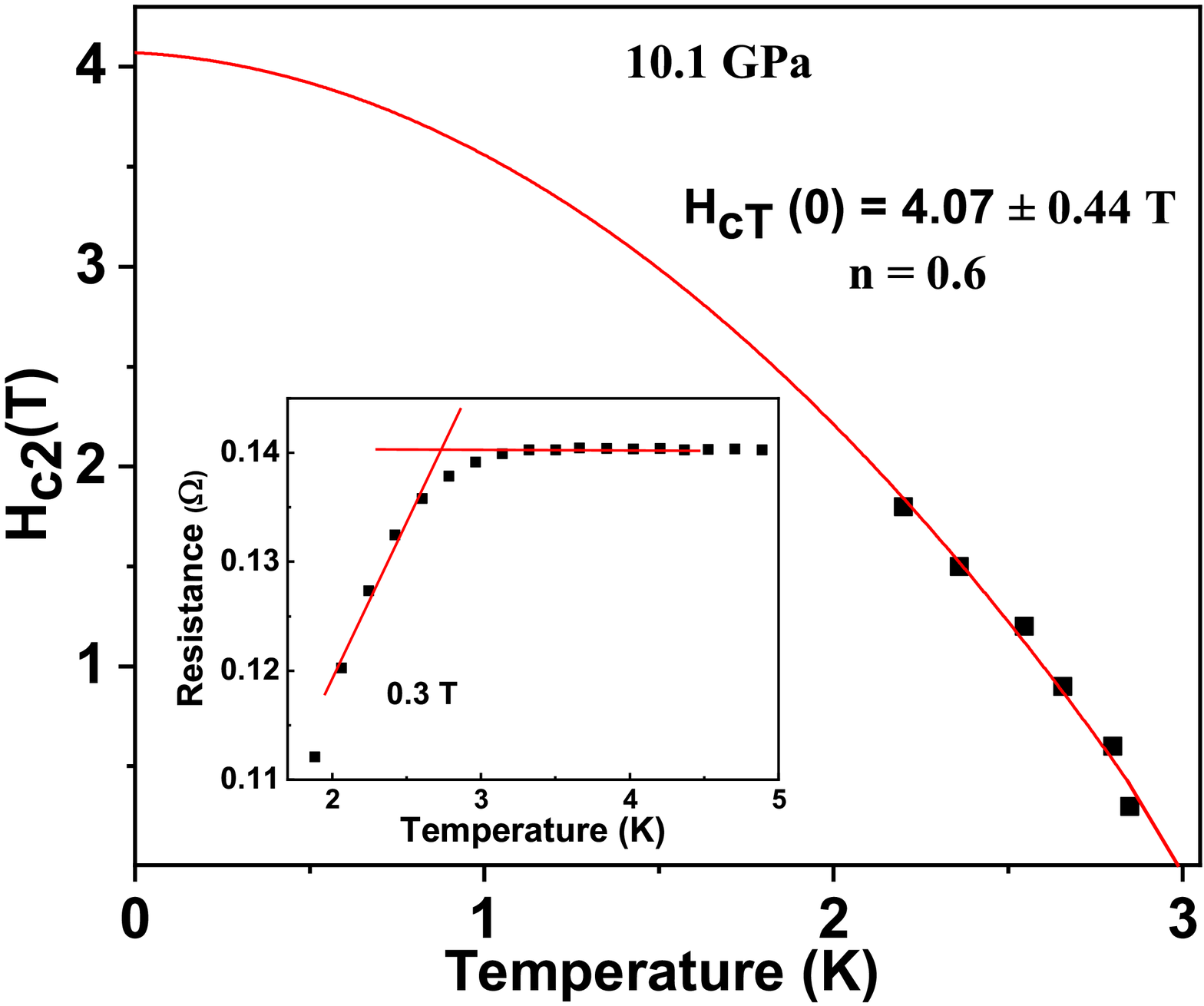}} \vspace{0.1 in}
\caption{Temperature dependence of upper critical field of TlInTe$_2$ at (a) 5.75 GPa (b) 10.1 GPa. The red line is extrapolated to obtain H$_{c2}$(0). The inset of figure b represents the procedure adopted to extract the critical temperature (T$_c$) from T vs resistance plot for different magnetic fields.}
\label{fig:Hc}
\end{figure}

\begin{figure}
    \centering
    \includegraphics[width=5.0in,height=3.5in]{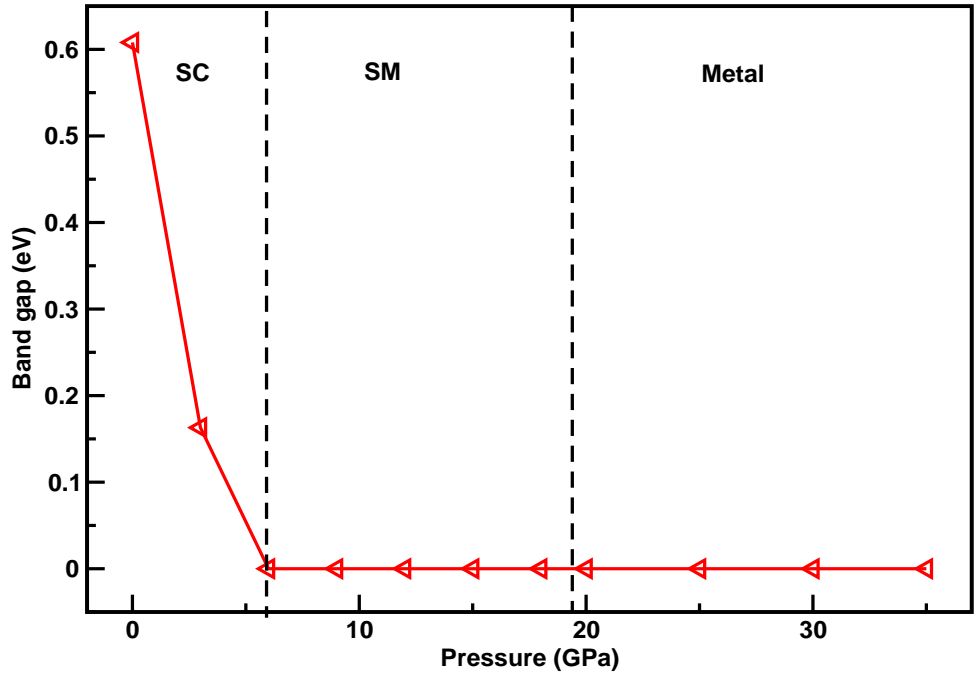}
    \caption{Calculated electronic band gap as a function of pressure for B37 phase without SOC.}
    \label{fig:Gap}
\end{figure}

\begin{figure}
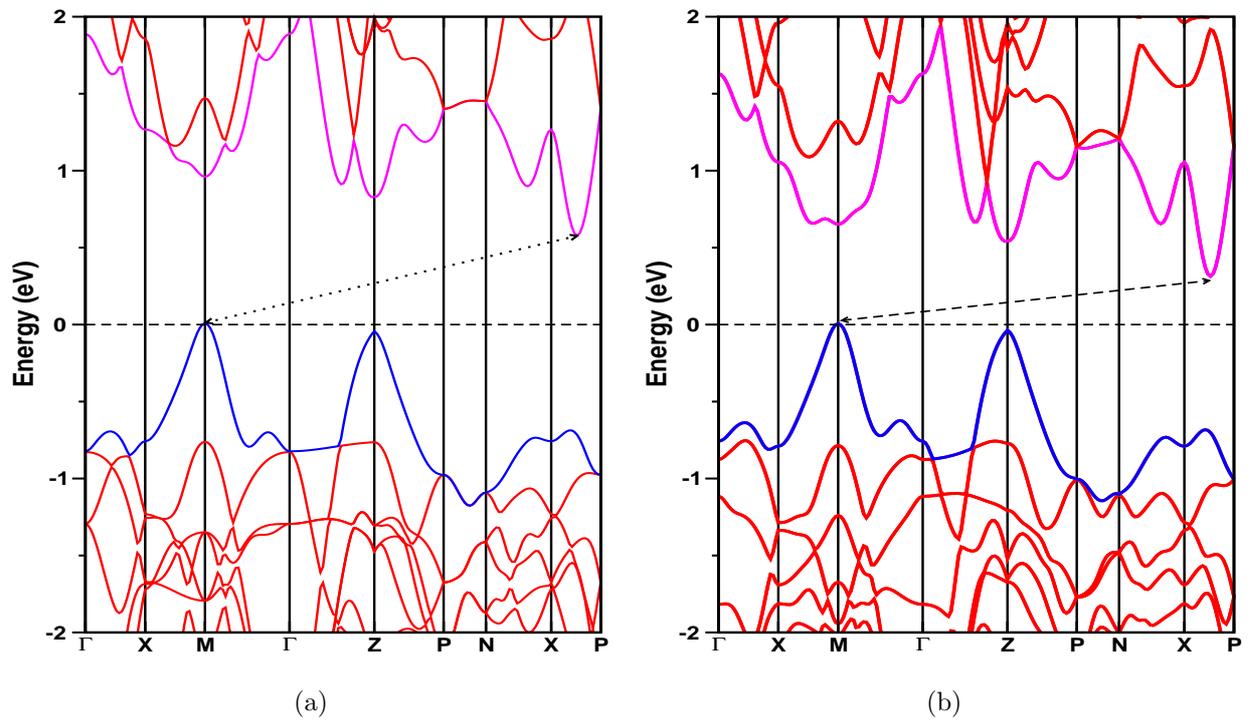

\centering
\subfigure[]{\includegraphics[width=3.1in,height=3.5in]{Figures/FigS10a.eps}} \hspace{2mm}
\subfigure[]{\includegraphics[width=3.1in,height=3.5in]{Figures/FigS10b.eps}} \vspace{0.1 in}
\caption{Calculated electronic band structure of B37 phase at ambient pressure (a) without and (b) with SOC.}
\label{fig:BS}
\end{figure}

\begin{figure}
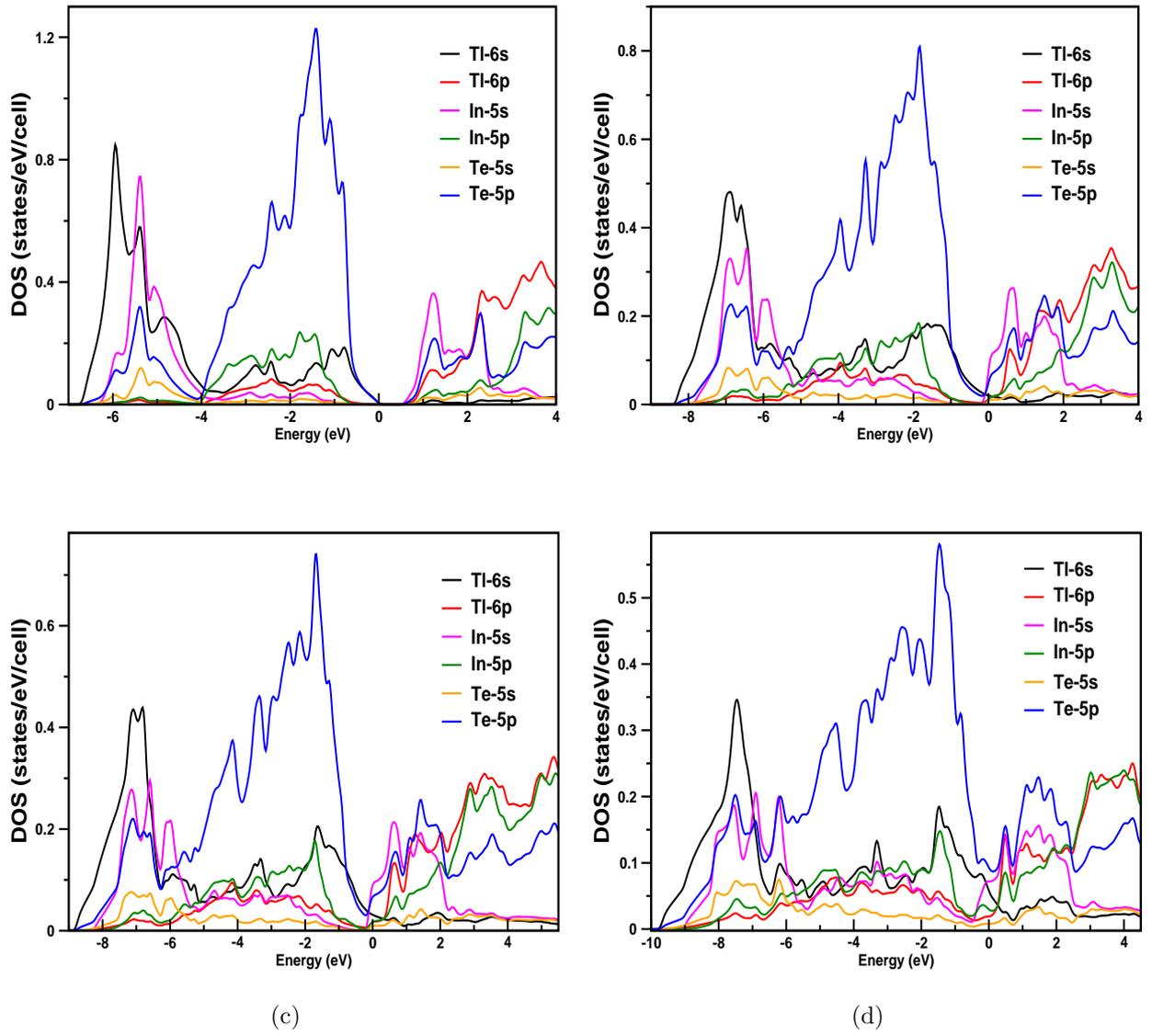

\centering
\subfigure[]{\includegraphics[width=3.1in,height=2.5in]{Figures/FigS11a.eps}} \hspace{2mm}
\subfigure[]{\includegraphics[width=3.1in,height=2.5in]{Figures/FigS11b.eps}} \vspace{0.1 in}
\subfigure[]{\includegraphics[width=3.1in,height=2.5in]{Figures/FigS11c.eps}} \hspace{2mm}
\subfigure[]{\includegraphics[width=3.1in,height=2.5in]{Figures/FigS11d.eps}}
\caption{Calculated projected density of states of B37 phase at (a) 0 GPa (b) 9 GPa (c) 15 GPa and (d) 30 GPa}
\label{fig:PDOS}
\end{figure}

\begin{figure}
\centering
\subfigure[]{\includegraphics[width=3.1in,height=2.8in]{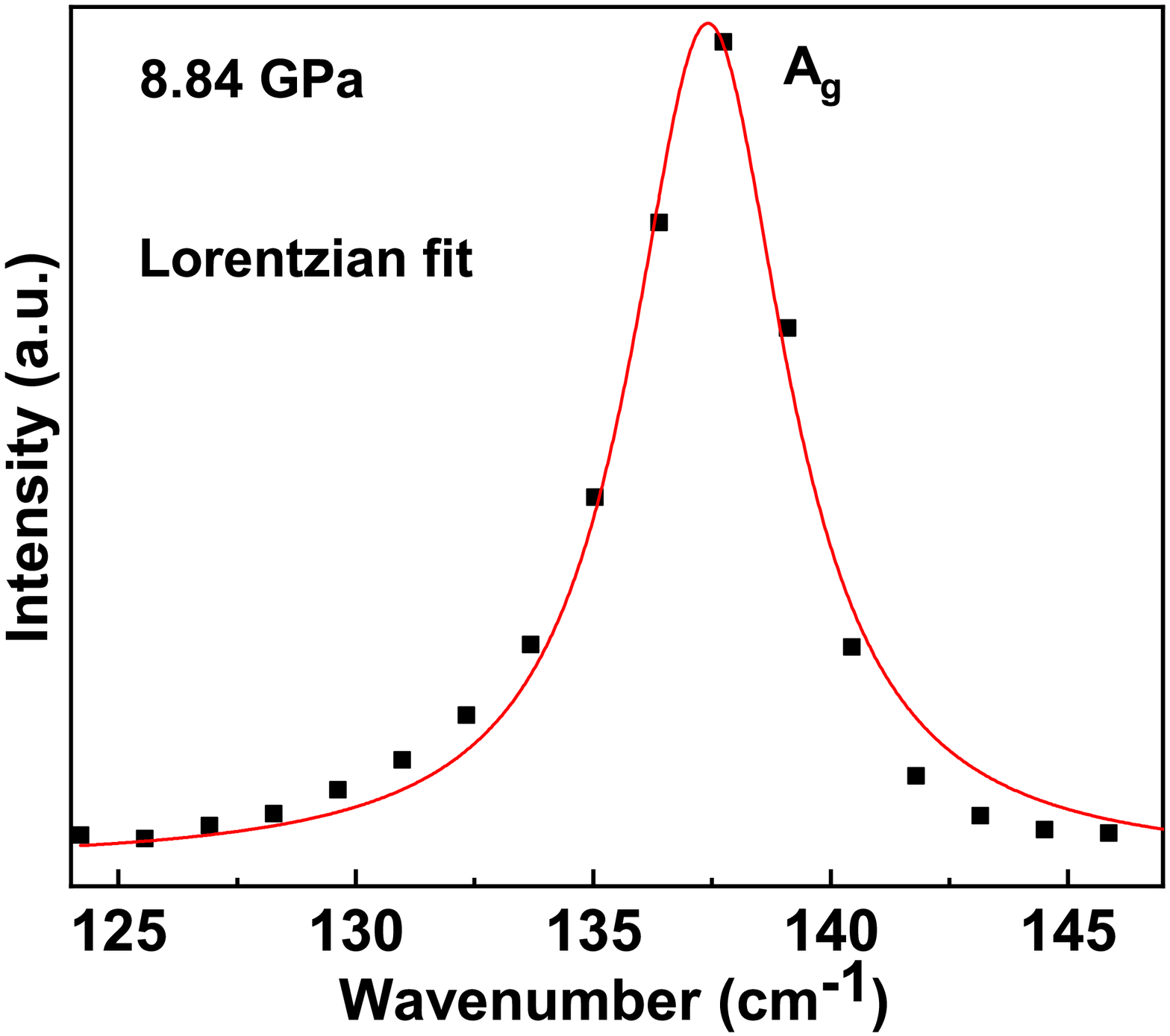}} \hspace{2mm}
\subfigure[]{\includegraphics[width=3.1in,height=2.8in]{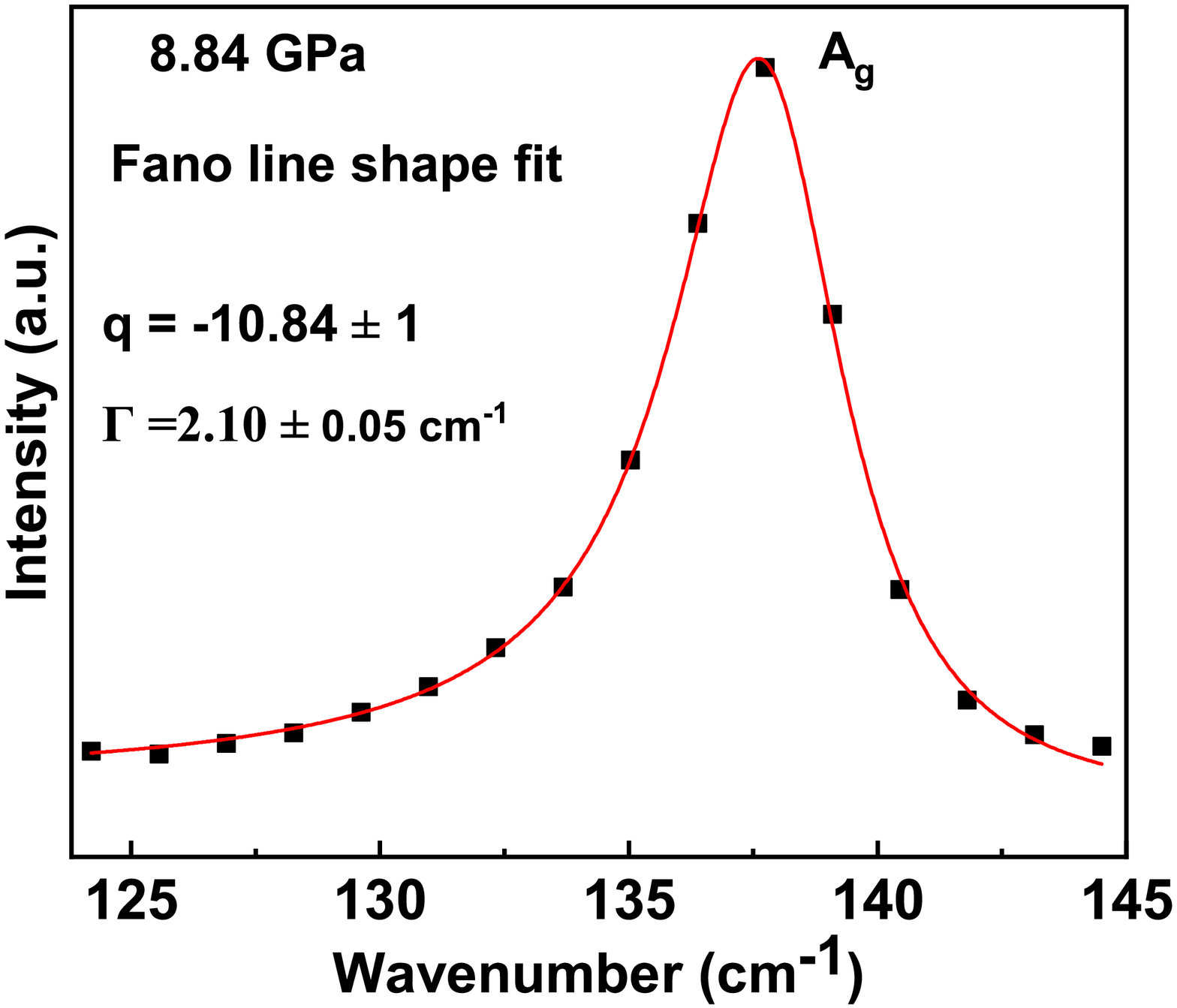}} \vspace{0.1 in}
\subfigure[]{\includegraphics[width=3.1in,height=2.8in]{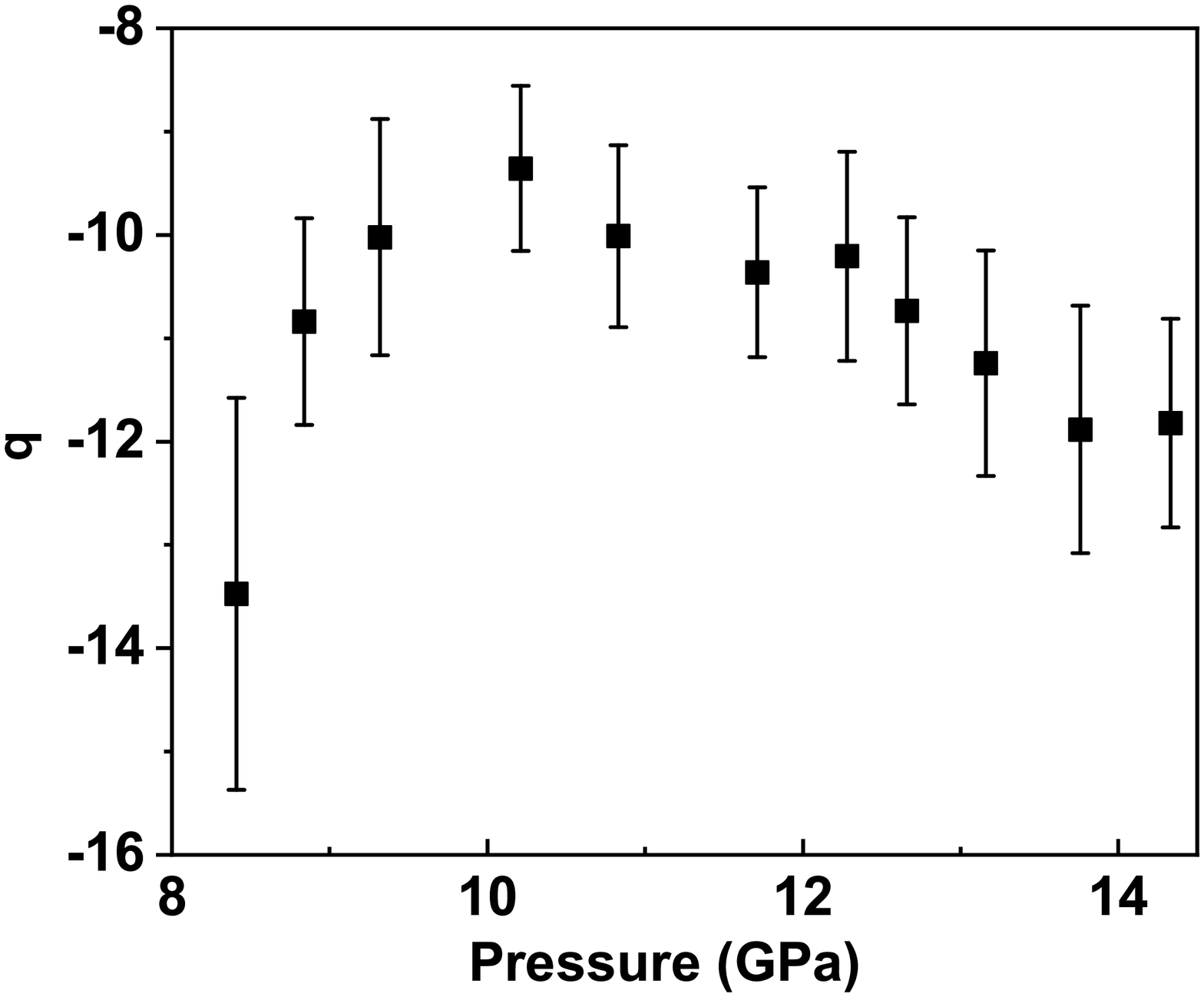}} \hspace{2mm}
\subfigure[]{\includegraphics[width=3.1in,height=2.8in]{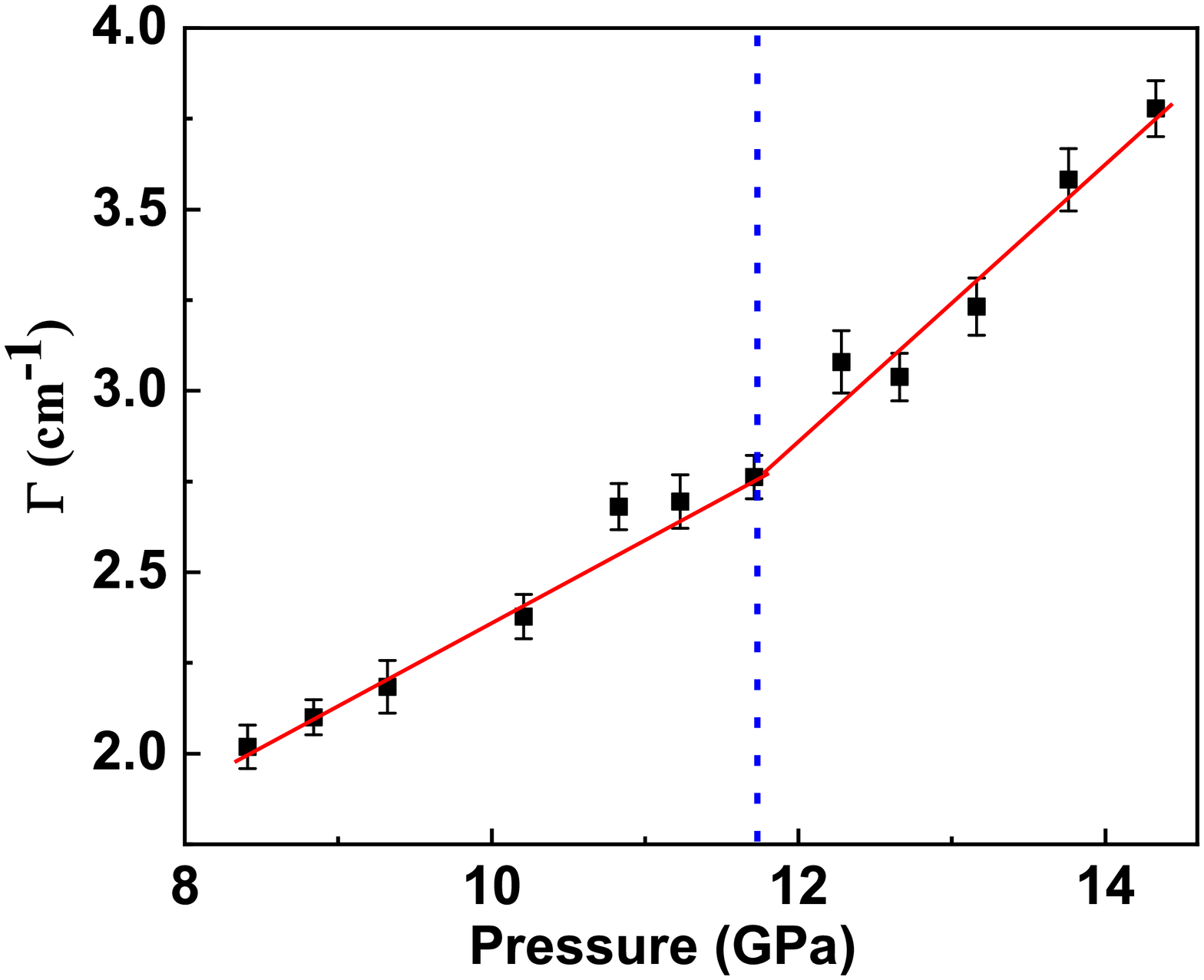}}
\caption{Raman spectrum of A$_g$ phonon mode of TlInTe$_2$ at 8.84 GPa fitted using (a) Lorentzian line shape (b) Fano line shape using equation 2 in the main text. (c) Pressure vs asymmetry parameter (q) of A$_g$ with error bars. (d) Pressure vs FWHM ($\Gamma$) of A$_g$ mode calculated using Fano line shape fit. The red line is the guidance to the eyes and the vertical dotted line is to show the slope change of phonon life time.}
\label{fig:Fanofit}
\end{figure}

\clearpage
\section{Bond parameters of B37 phase under high pressure}
We have also analyzed the bond lengths and bond angles from the relaxed structures for the ambient phase to get more insight on crystal structure of B37 phase under pressure [Fig. S13 and S14]. The bond lengths of TlInTe$_2$ do not exhibit any appreciable change except for the subtle anomalies of In-Te, Tl-Tl and In-In bond lengths around 25 GPa. As revealed in Fig. S14, we have observed discontinuities in the calculated in-equivalent Te-Tl-Te bond angles of distorted Thomson cube of Tl-Te$_8$ at 5-6 GPa and 10-12 GPa. The former discontinuity in the bond angles is due to semi conductor to semi metal transition whereas the later is correlated to the giant A$_g$ mode softening.

\begin{figure}
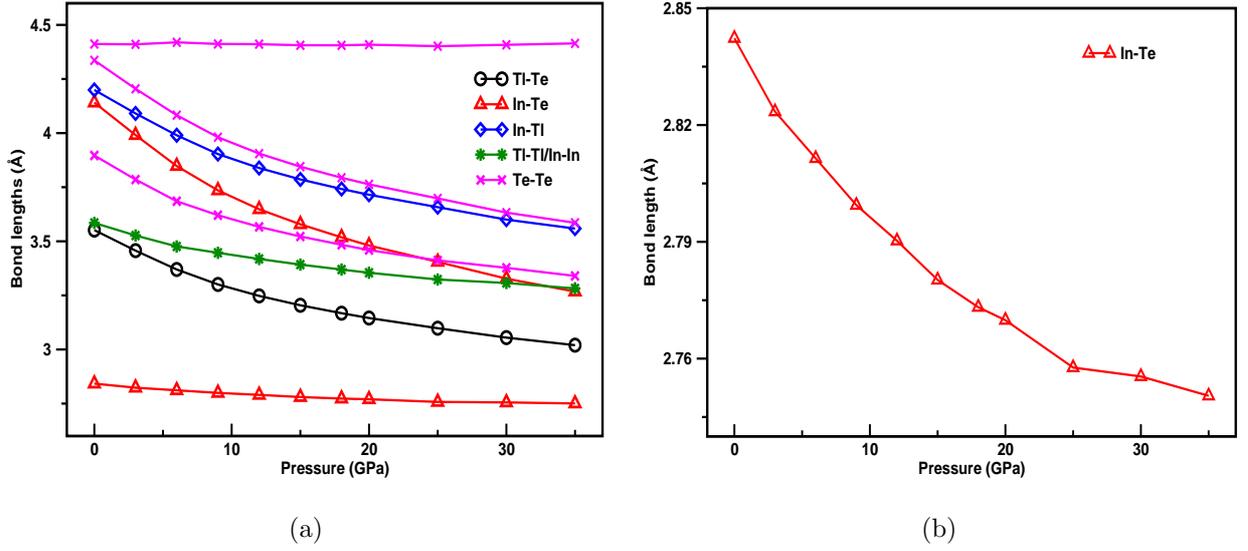

\centering
\subfigure[]{\includegraphics[width=3.1in,height=2.5in]{Figures/FigS13a.eps}} \hspace{2mm}
\subfigure[]{\includegraphics[width=3.1in,height=2.5in]{Figures/FigS13b.eps}} \vspace{0.1 in}
\caption{Calculated (a) bond lengths (b) In-Te bond length of B37 phase of TlInTe$_2$ as a function of pressure.}
\label{fig:bondlength}
\end{figure}

\begin{figure}
    \centering
    \includegraphics[width=6.8in,height=5.3in]{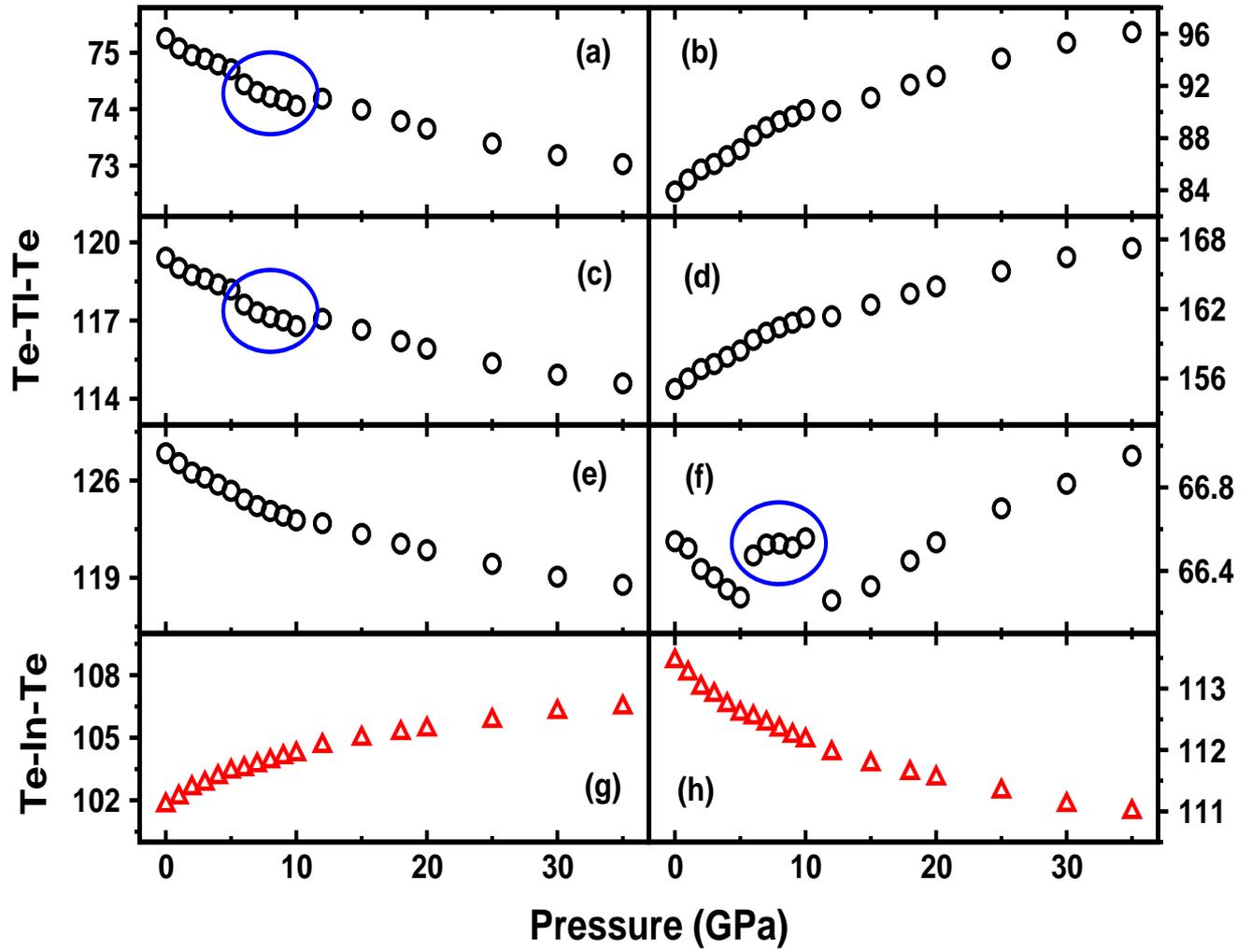}
    \caption{Calculated bond angles of B37 phase of TlInTe$_2$ as a function of pressure. (a-f) the in-equivalent bond angles of Thomson cube Tl-Te$_8$ (open circles in black) and (g,h) the in-equivalent bond angles of tetrahedra In-Te$_4$ (open upper triangles in red). The discontinuities in the in-equivalent bond angles of Thomson cube (Tl-Te$_8$) (highlighted as blue circle) at 6 GPa is due to semi conductor to semi metal transition whereas the later one between 10-12 GPa is correlated to the giant A$_g$ mode softening.}
    \label{fig:BA}
\end{figure}
\begin{figure}
    \centering
    \includegraphics[width=6.8in,height=4.0in]{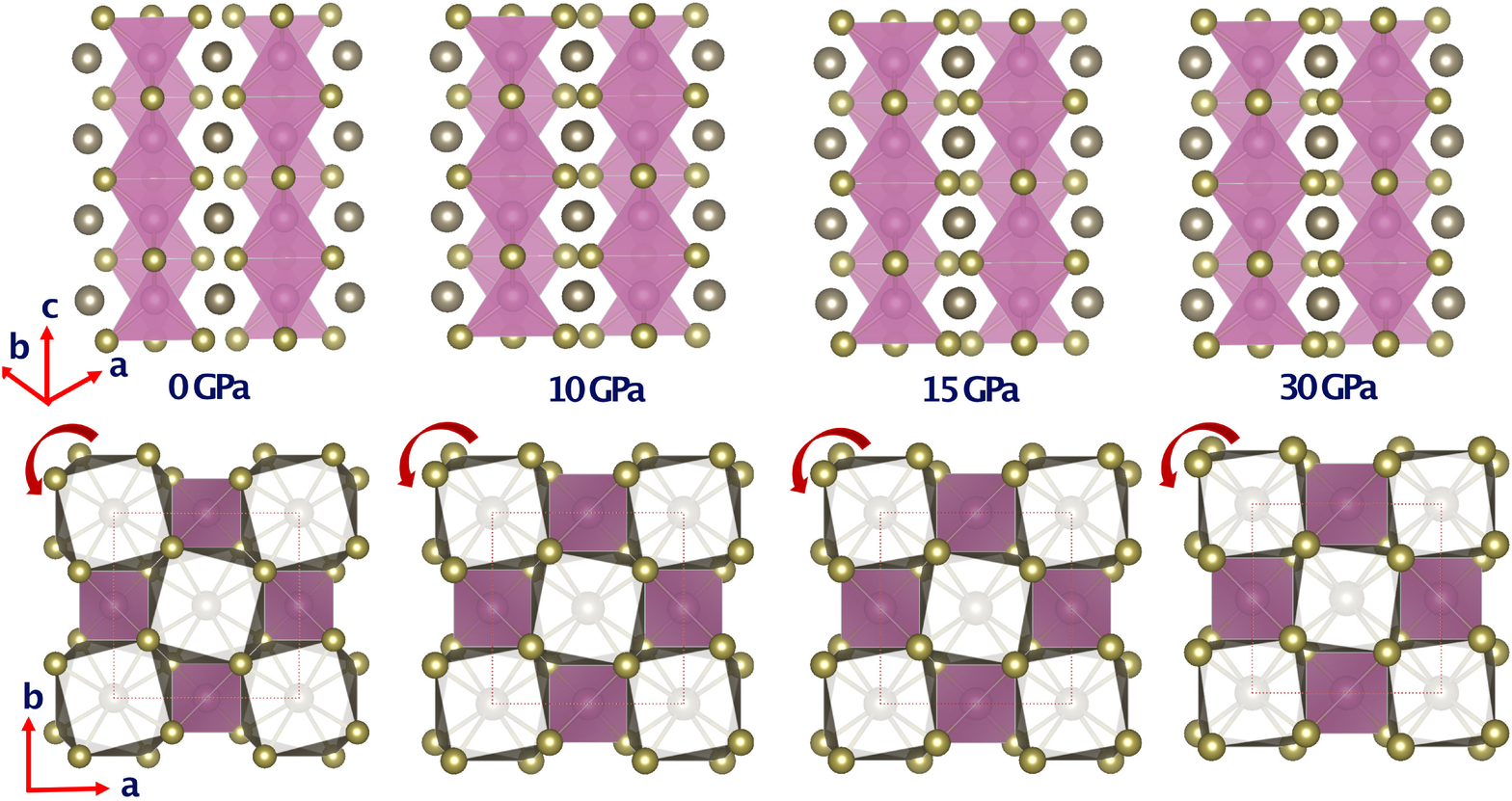}
    \caption{Rattler behavior of Tl$^{+}$ (top) and rotation of top and bottom square planers of adjacent Thompson cubes (bottom) of B37 phase as a function of pressure viewing from the $ab$-plane.}
    \label{fig:TCR}
\end{figure}

\clearpage
\bibliographystyle{plain}
\bibliography{references}